\documentclass[11pt]{article}
\PassOptionsToPackage{hyperfootnotes=false}{hyperref}
\usepackage[preprint]{acl}
\usepackage{times}
\usepackage{latexsym}
\usepackage[T1]{fontenc}
\usepackage[utf8]{inputenc}
\usepackage{microtype}
\usepackage{inconsolata}
\usepackage{graphicx}
\usepackage{placeins}
\usepackage{pifont}
\usepackage{amsmath,amssymb,booktabs,array,tabularx,tikz}
\usetikzlibrary{arrows.meta,positioning,shapes.geometric,calc}
\newcolumntype{L}[1]{>{\raggedright\arraybackslash}p{#1}}
\newcolumntype{Y}{>{\raggedright\arraybackslash}X}
\newcommand{\bench}{MESH}

\title{MESH: Multi-receiver Evaluation of Spatial Hearing \\
across Egocentric and Allocentric Views}
\author{Wen Wen\textsuperscript{1}\thanks{Equal contribution.} \quad
Bohan Li\textsuperscript{1*} \quad
KwanHo Tam\textsuperscript{1} \quad Yu Xi\textsuperscript{1} \\
\textbf{Jing Peng}\textsuperscript{1} \quad \textbf{Shuai Wang}\textsuperscript{2} \quad \textbf{Kai Yu}\textsuperscript{1} \\
\textsuperscript{1}Shanghai Jiao Tong University \quad
\textsuperscript{2}Nanjing University \\
\texttt{wenyideng@sjtu.edu.cn, everlastingnight@sjtu.edu.cn}}
\makeatletter
\def\input@path{{manuscript/}}
\makeatother
\newcommand{\taskDirectionEstimation}[1][ ]{\mbox{ES-Dir}}
\newcommand{\taskDistanceEstimation}[1][ ]{\mbox{ES-Dist}}
\newcommand{\taskTextDirection}[1][ ]{\mbox{ES-CDir}}
\newcommand{\taskTextDistance}[1][ ]{\mbox{ES-CDist}}
\newcommand{\taskEgoStaticASR}[1][ ]{\mbox{ES-Trans}}
\newcommand{\taskSourceLocalization}[1][ ]{\mbox{AS-Pos}}
\newcommand{\taskTextLocalization}[1][ ]{\mbox{AS-CPos}}
\newcommand{\taskGlobalStaticASR}[1][ ]{\mbox{AS-Trans}}
\newcommand{\taskDirectionTracking}[1][ ]{\mbox{ED-Dir}}
\newcommand{\taskDistanceTracking}[1][ ]{\mbox{ED-Dist}}
\newcommand{\taskEgoDynamicASR}[1][ ]{\mbox{ED-Trans}}
\newcommand{\taskSourceTracking}[1][ ]{\mbox{AD-CPos}}
\newcommand{\taskGlobalDynamicASR}[1][ ]{\mbox{AD-Trans}}

\begin{document}
\maketitle

\begin{abstract}
Spatial hearing in multi-receiver environments requires identifying who
speaks, where they are, and how they move.
Existing benchmarks evaluate spatial prediction, transcription,
tracking, and distributed speech processing, but offer limited
evaluation of whether speech signal remains associated with the correct
physical source across receivers and motion.
We introduce \bench{}, which evaluates this association through
egocentric and allocentric views of shared acoustic scenes.
The egocentric view uses local recordings and peer metadata,
while the allocentric view combines synchronized recordings from
multiple receivers in a shared 3D world frame.
\bench{} contains nearly 40,000 questions across 13 static and
dynamic localization and spatial transcription tasks.
We evaluate eleven audio-model configurations and a language-model
agent using specialist tools.
The best static distance-estimation and dynamic direction-tracking
accuracies are only 21.90\% and 41.54\%, respectively, while the
lowest word error rates across four transcription tasks range from
59.15\% to 73.10\%.
Ablation studies and additional experiments reveal that existing models are not yet able to effectively leverage peer information and struggle to handle multi-speaker and dynamic scenarios.
These findings highlight the gap between access to distributed acoustic
information and its effective use for spatial speech understanding.
We release the dataset and evaluation code to support further
research on collaborative and distributed spatial hearing.\footnote{
Evaluation code:
\url{https://github.com/Wenanzhi/MESH-Benchmark}.
Dataset:
\url{https://huggingface.co/datasets/Heiheihaha17/MESH-Benchmark}.
}
\end{abstract}

\section{Introduction}
Real acoustic environments contain multiple sound sources, some stationary and others moving, that jointly form a complex and changing acoustic scene. Their emissions overlap with environmental noise and reverberation, making it difficult to determine what each speaker says, where they are, and how they move. We consider a distributed listening setting in which multiple receivers equipped with microphone arrays observe this scene from different locations. In embodied applications, these receivers may also move and emit sound, making them both observers and participants in the acoustic environment. Each receiver captures spatial evidence from its own position, while shared receiver poses, known playback signals, and recordings from other locations provide additional information about the same scene. This setting motivates two coordinate-based views: understanding surrounding sources in an individual receiver's local frame, and understanding source locations and speech in a shared 3D world frame. Peer collaboration and distributed recordings provide information for these views. An exhibition hall provides a concrete example: guide robots operate among visitors who stand near exhibits or walk between displays, while visitors and robots may speak simultaneously. A listening system must associate each utterance with the correct visitor, locate that visitor, and follow their motion as the relative geometry changes.

Existing benchmarks evaluate important components of this problem, including numerical spatial prediction and spatially selective transcription \citep{liu2026jaeger,zhu2026spatialomni,kumar2026sealm}, while distributed-recording resources support speech processing across devices \citep{schmalenstroeer2023libriwasn}. However, performance on an individual task does not establish that a system can correctly associate speech, location, and motion in a shared scene: a correct direction estimate may leave distance unresolved, and a fluent transcript may belong to the wrong speaker. We introduce \textbf{\bench{}}: \textbf{M}ulti-receiver \textbf{E}valuation of \textbf{S}patial \textbf{H}earing across Egocentric and Allocentric Views. The egocentric view is a receiver's self view: spatial queries and predictions use its local coordinate frame. The allocentric view instead uses a shared 3D world coordinate system. We define these views by their coordinate frames and specify their information inputs separately. In our evaluation, egocentric tasks use one receiver's recording with peer collaboration supplying robot poses, playback transcripts, and playback intervals; allocentric tasks use all available receiver recordings and metadata. These input protocols leave the system architecture open. An audio-language model may predict answers directly, whereas an agentic system may coordinate signal-processing tools and specialist models to obtain and combine perceptual evidence \citep{wang2026audiomind}. Both must meet the same task requirements: a supplied transcript identifies a source to localize, a spatial cue identifies speech to transcribe, and a scene-level query requests source locations or trajectories.

We instantiate this setting with approximately 40,000 questions across eight static and five dynamic tasks. The dataset contains 800 base scenes spanning 100 rooms; each base scene produces one static and two independently sampled dynamic realizations that share speech content, noise, and initial layout. Synchronized first-order Ambisonic recordings capture spatial observations at each receiver, while prescribed source and receiver motion changes the geometry over time. Static and dynamic localization and spatial transcription are therefore evaluated within shared scenes using physical errors and target-specific word error rates. We evaluate eleven audio-model configurations and an agentic reference system. The best static distance-estimation accuracy is 21.90\% within 20\% relative error, and the best direction-tracking accuracy is 41.54\% within $20^\circ$. The lowest word error rates across the four transcription tasks remain 59.15--73.10\%. Analyses across source counts, acoustic conditions, and motion, supported by original model responses, reveal wrong-speaker transcription, inaccurate distance estimates, and nearly constant predictions despite changing source--receiver distances. These results identify limitations in the basic perception needed for scene understanding and motivate further study of how peer information and distributed observations can support reliable associations between speech and its physical source.
\section{Related Work}
\label{sec:related-work}

\paragraph{Spatial audio evaluation.}
AIR-Bench, MMAU, and AudioBench evaluate broad audio-language capabilities
\citep{yang-etal-2024-air,sakshi2025mmau,wang-etal-2025-audiobench}.
STAR-Bench assesses acoustic perception and spatio-temporal reasoning
\citep{liu2026star}, while RMS-AQA couples event grounding with spatial
and temporal multiple-choice reasoning, including moving-source scenes
\citep{chen2026rmsaqa}.
SpatialSoundQA and BiDepth use discretized spatial answers
\citep{zheng2024bat,biswas2025owl}, whereas SpatialSceneQA, SO-Bench,
and SEA-LM also include numerical spatial prediction
\citep{liu2026jaeger,zhu2026spatialomni,kumar2026sealm}.
ST-AudioQA evaluates trajectories in its encoder, while its language
model primarily answers categorical, binned, and relational questions
\citep{oh2026staudio}; we mark its trajectory coverage as partial.
Table~\ref{tab:benchmark_comparison} compares numerical localization
and trajectories, spatially cued transcription, joint multi-receiver
input, and receiver motion.
Full numerical coverage requires direct spatial estimates scored by
physical error, rather than multiple-choice answers or coarse bins.

\begin{table*}[t]
\centering
\small
\setlength{\tabcolsep}{2pt}
\renewcommand{\arraystretch}{1.12}
\DeclareRobustCommand{\partialcoverage}{\textcolor{orange!85!black}{\tikz[baseline=-0.55ex]{%
  \fill (0,0) -- (0,0.65ex) arc[start angle=90,end angle=270,radius=0.65ex] -- cycle;
  \draw[line width=0.35pt] (0,0) circle[radius=0.65ex];}}}
\DeclareRobustCommand{\covered}{\textcolor{green!55!black}{\ding{51}}}
\DeclareRobustCommand{\notcovered}{\textcolor{red!80!black}{\ding{55}}}
\newcommand{\comparisonheader}[1]{\textbf{\shortstack{#1}}}
\begin{tabularx}{\textwidth}{@{}Y*{5}{c}@{}}
\toprule
\textbf{Benchmark}
& \comparisonheader{Numerical spatial\\localization}
& \comparisonheader{Numerical\\trajectories}
& \comparisonheader{Spatially cued\\transcription}
& \comparisonheader{Joint multi-receiver\\input}
& \comparisonheader{Receiver\\motion} \\
\midrule
\mbox{SpatialSoundQA \citep{zheng2024bat}}
& \partialcoverage & \notcovered & \notcovered & \notcovered & \notcovered \\
\mbox{BiDepth \citep{biswas2025owl}}
& \partialcoverage & \notcovered & \notcovered & \notcovered & \notcovered \\
\mbox{STAR-Bench \citep{liu2026star}}
& \notcovered & \notcovered & \notcovered & \notcovered & \notcovered \\
\mbox{SpatialSceneQA \citep{liu2026jaeger}}
& \covered & \notcovered & \notcovered & \notcovered & \notcovered \\
\mbox{ST-AudioQA \citep{oh2026staudio}}
& \partialcoverage & \partialcoverage & \notcovered & \notcovered & \notcovered \\
\mbox{SO-Bench \citep{zhu2026spatialomni}}
& \covered & \notcovered & \covered & \notcovered & \notcovered \\
\mbox{SEA-LM Eval \citep{kumar2026sealm}}
& \covered & \notcovered & \covered & \notcovered & \notcovered \\
\mbox{RMS-AQA \citep{chen2026rmsaqa}}
& \partialcoverage & \notcovered & \notcovered & \notcovered & \notcovered \\
\mbox{LOCATA \citep{evers2020locata}}
& \covered & \covered & \notcovered & \notcovered & \covered \\
\mbox{L3DAS22 \citep{guizzo2022l3das22}}
& \covered & \notcovered & \notcovered & \covered & \notcovered \\
\midrule
\textbf{\bench{}}
& \covered & \covered & \covered & \covered & \covered \\
\bottomrule
\end{tabularx}
\caption[Comparison of evaluation protocols for spatial audio]{Comparison of spatial audio evaluation protocols.
\covered: explicitly evaluated;
\partialcoverage: discretized output or auxiliary evaluation;
\notcovered: not evaluated in the public protocol.
Comparison criteria are defined in Section~\ref{sec:related-work}.}
\label{tab:benchmark_comparison}
\end{table*}

\paragraph{Distributed perception and target speech selection.}
LOCATA evaluates localization and tracking with source and array motion,
reporting results for each array separately \citep{evers2020locata}.
L3DAS22 evaluates spatial speech enhancement and localization
\citep{guizzo2022l3das22}, while LibriWASN and CHiME-7 support speech
processing across distributed devices
\citep{schmalenstroeer2023libriwasn,cornell2023chime7}.
DSS LibriSpeech evaluates localization-informed transcription
\citep{tang2024spatialaudio}, and DAE-TSE uses partial transcripts to
cue target speaker extraction \citep{li2026dae}.
Here, joint multi-receiver input requires joint use of recordings from
spatially separated devices observing the same scene, not merely
multiple audio channels; receiver motion refers to movement of the
recording device itself.
\bench{} combines corresponding receiver views with shared-base static
and dynamic scenes, using transcripts to identify speakers for numerical
localization and spatial cues to specify whose complete input speech
to transcribe.
\section{MESH benchmark}
\label{sec:benchmark}

\bench{} evaluates spatial speech understanding from two coordinate-based
views: the \textbf{egocentric view} expresses spatial queries and predictions
in a receiver's local frame, while the \textbf{allocentric view} uses a shared
3D world coordinate system.
Both views address the same question: \emph{who said what and where?}
We define the views by their spatial reference frames and separately specify
the recordings and collaborative information available in our evaluation.

\begin{figure*}[!t]
\centering
\includegraphics[width=\textwidth]{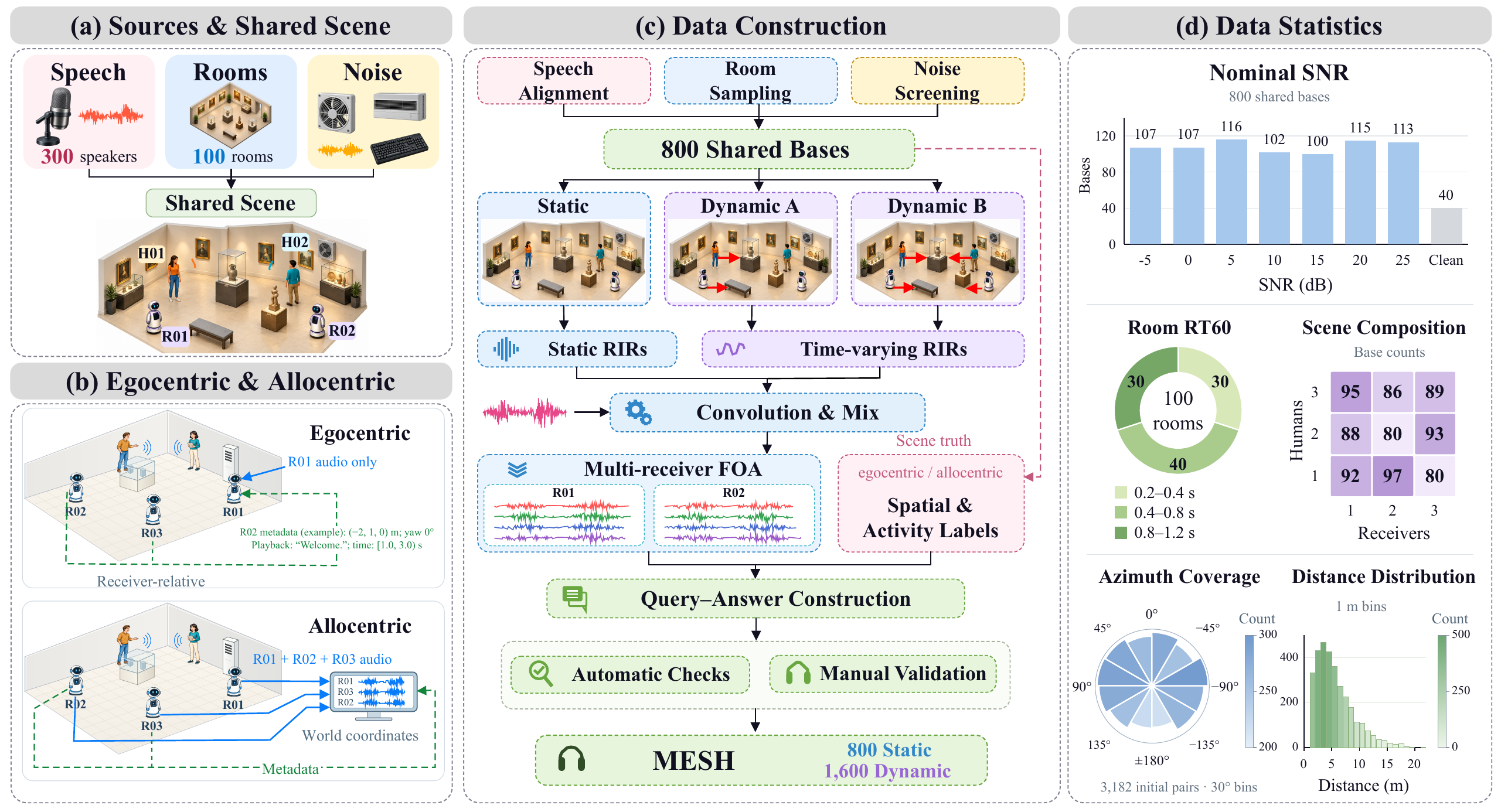}
\caption{Overview of \bench{}: (a) scene sources, (b) coordinate views and available information, (c) construction and validation workflow, and (d) acoustic and spatial coverage. Panels are explained in Section~\ref{sec:benchmark}; scene drawings and metadata examples are schematic.}
\label{fig:construction}
\vspace{-10pt}
\end{figure*}

We organize the benchmark around two task objectives:
\textbf{localization}, which estimates where sound sources are,
and \textbf{transcription}, which recovers what a spatially
specified person said.
Each objective is evaluated under applicable \textbf{static} and
\textbf{dynamic} motion settings.
Static tasks concern fixed source and receiver positions, while
dynamic tasks require localization or transcription as spatial
relationships change over time.
Together, the two views and motion settings define 13 tasks.
Figure~\ref{fig:construction} summarizes the benchmark: panel (a) combines
speech, rooms, and noise into a shared scene; panel (b) shows the two
coordinate views and their information inputs; panel (c) constructs
recordings, labels, and questions; panel (d) characterizes the resulting
acoustic conditions and spatial coverage.
Table~\ref{tab:task-overview} in Appendix specifies the input cues and outputs for all
tasks under both views.

\subsection{Scene and Coordinate Views}
\label{sec:observations}

\paragraph{Scene setup.}
Each scene contains $H\in\{1,2,3\}$ humans and
$N\in\{1,2,3\}$ robot receivers.
The figure uses H01, H02, etc. for humans and R01, R02, etc. for robot
receivers. Robot $i$ records a four-channel first-order Ambisonic (FOA)
signal $X_i\in\mathbb{R}^{4\times L}$, where $L$ is the number
of samples per channel.
FOA encodes omnidirectional pressure and three directional components
at one receiver location. All receivers record over the same time interval.
Speech sources include humans and robots playing speech.
For each source $s$, $c_s$ denotes its transcript, and
$\mathbf{x}_s(t)\in\mathbb{R}^3$ denotes its world-frame
position at time $t$.
The known pose of robot $i$ is
$P_i(t)=(R_i(t),\mathbf{x}_i(t))$,
where $\mathbf{x}_i(t)$ is the world-frame position of its
local origin and $R_i(t)\in\mathbb{R}^{3\times3}$ is the
rotation matrix describing its orientation.
A point with robot-local coordinates $\mathbf{r}$ therefore
has world coordinates $R_i(t)\mathbf{r}+\mathbf{x}_i(t)$.

\paragraph{Egocentric view and peer collaboration.}
The egocentric view is robot $i$'s \emph{self view}: source locations and
spatial queries are expressed in its local coordinate frame, for example
as a speaker's direction or distance relative to the robot.
The relative position of source $s$ is
\begin{equation}
\mathbf{r}_{is}(t)
=
R_i(t)^\top
\left[\mathbf{x}_s(t)-\mathbf{x}_i(t)\right].
\label{eq:spatial-perspectives}
\end{equation}

In our evaluation, robot $i$ observes its own recording $X_i$.
Peer collaboration supplies robot metadata
$M_j=(P_j,c_j^{\mathrm{out}},\tau_j)$,
containing robot $j$'s pose, speech playback transcript,
and playback intervals.
This information helps distinguish human speech from robot playback
without sharing peer recordings or human speech annotations.
Given public room context $K$ and a query $q_i$, the prediction is
\begin{equation}
\hat{y}_i^{\mathrm{ego}}
=
f_\theta^{\mathrm{ego}}
\left(X_i,\{M_j\}_{j=1}^{N},K,q_i\right).
\label{eq:egocentric-inference}
\end{equation}
All spatial cues and predictions use robot $i$'s local frame.

\paragraph{Allocentric view.}
The allocentric view represents source locations in a shared 3D world
coordinate system as $\mathbf{x}_s(t)$, independent of any particular
receiver's position or orientation.
In our evaluation, it receives synchronized recordings and metadata
from all available receivers.
Given an allocentric query $q^{\mathrm{allo}}$, the prediction is
\begin{equation}
\hat{y}^{\mathrm{allo}}
=
f_\theta^{\mathrm{allo}}
\left(
\{X_i,M_i\}_{i=1}^{N},K,q^{\mathrm{allo}}
\right).
\label{eq:global-inference}
\end{equation}
The system must associate acoustic observations across receivers
with the same physical source.
When $N=1$, the two views use the same recording and robot metadata,
but spatial cues and outputs remain in the receiver-local and world frames,
respectively. The coordinate frame defines the view; the input protocol
specifies the available information and leaves model architectures open.

\paragraph{Motion settings and queries.}
Static scenes keep source and receiver poses fixed.
In dynamic scenes, humans, robots, or both may move.
Source motion changes its world position, while receiver motion
changes the observed direction and distance even for a stationary source.

Each task provides a query specifying the target or requested output.
Localization queries may ask for all sources or use speech content
or spatial cues to select a target.
Transcription queries use spatial cues to select whose speech
to recover.
Corresponding queries can refer to the same physical source across
receivers and views, with spatial cues expressed in the
appropriate coordinate frame.
Only annotations explicitly provided as query cues are visible;
other human annotations are reserved for evaluation.
Questions are evaluated independently, without scene images
or access to other tasks' questions and answers.

\paragraph{Task identifiers.}
The codes in Table~\ref{tab:task-reference} specify view
(E: egocentric; A: allocentric), motion setting (S: static; D: dynamic),
and output (Dir: direction; Dist: distance; Pos: world position;
Trans: transcription). An optional C marks a supplied transcript
that selects the source for localization. Thus, ES-CDir is
static egocentric transcript-conditioned direction localization.
Dynamic localization returns a trajectory of the named quantity;
tracking is not a separate objective. ED-Dist uses an initial
spatial cue, not a transcript, so it has no C.

\subsection{Egocentric Tasks}
\label{sec:egocentric-tasks}

Egocentric tasks express spatial cues and predictions in the observing
robot's local coordinate frame. They use the robot's own recording
with peer metadata supplying additional spatial and speech information.

\paragraph{Static localization.}
\emph{Direction localization} (\taskDirectionEstimation{}) estimates the azimuth and elevation
of all humans and other robots producing audible speech playback,
given the source count.
\emph{Distance localization} (\taskDistanceEstimation{}) selects one human through a unique
azimuth sector and estimates that person's distance.
\emph{Transcript-conditioned direction localization} (\taskTextDirection{}) and \emph{Transcript-conditioned distance localization} (\taskTextDistance{}) instead use a supplied
speech transcript to identify a human and estimate the speaker's
direction or distance, respectively.
The latter two tasks share the same transcript, target, and receiver,
testing two spatial properties of a speaker identified by content.

\paragraph{Static transcription.}
\emph{Spatially cued transcription} (\taskEgoStaticASR{}) uses a person's direction and distance
to select whose speech to transcribe.
The system must recover all speech produced by that person
during the input clip.
Unlike localization, which predicts where a source is,
transcription uses location to select the speech to recover.

\paragraph{Dynamic localization.}
Given H, the number of humans, direction trajectory localization (ED-Dir) estimates the azimuth and elevation of each person over time while maintaining consistent anonymous identities. Audio played by the robot is treated as interference. For distance trajectory localization (ED-Dist), a target person is identified using an initial speech interval and azimuth sector. The task is then to track that person's distance over time, even after they move outside the initial sector. Both tasks require consistent source identification as the spatial positions of the speakers change.
\emph{Distance trajectory localization} (\taskDistanceTracking{}) uses an initial speech interval and
azimuth sector to identify one human, then estimates that
person's distance over time, even after they leave the sector.
Both tasks require maintaining the same source identity
as spatial observations change.

\paragraph{Dynamic transcription.}
\emph{Spatially cued transcription} (\taskEgoDynamicASR{}) provides two speech-active spatial anchors
at least 1~s apart, each specifying direction, distance, or both.
The anchors identify the same person, whose complete speech
must be transcribed throughout the clip, including speech
outside the anchor times.

\subsection{Allocentric Tasks}
\label{sec:global-tasks}

Allocentric tasks express spatial cues and predictions in a shared 3D
world coordinate system and use all available receiver recordings.
They require associating acoustic observations across receivers
and expressing spatial predictions in world coordinates.

\paragraph{Static localization.}
Given the human count $H$, \emph{source-set position localization} (\taskSourceLocalization{}) estimates
the world-frame positions of all humans as an unordered set
of $H$ points.
\emph{Transcript-conditioned position localization} (\taskTextLocalization{}) instead uses a supplied speech transcript
to identify one human and estimate that person's world position.
Both tasks require three-dimensional localization, but differ
in whether the target is selected by speech content.

\paragraph{Static transcription.}
\emph{Region-cued transcription} (\taskGlobalStaticASR{}) selects a human through a supplied
region in world coordinates and asks for that person's complete
speech.
If the region contains no human, the system must return
an empty transcript.

\paragraph{Dynamic localization.}
\emph{Transcript-conditioned trajectory localization} (\taskSourceTracking{}) uses a supplied speech transcript
to identify one human and requests that person's
world-coordinate trajectory.
The system must maintain the same source association
as the person moves and acoustic observations change.

\paragraph{Dynamic transcription.}
\emph{Trajectory-cued transcription} (\taskGlobalDynamicASR{}) provides the target human's complete
world trajectory, including silent periods but without
speech-activity labels.
The system must use the trajectory to identify the person
across receiver recordings and transcribe all their speech
throughout the clip.

\subsection{Data Curation Pipeline}
\label{sec:data}
We combine recorded speech and environmental noise with acoustic simulation
to obtain consistent source identities and spatial annotations across
receivers and motion settings (Figure~\ref{fig:construction}).

\paragraph{Shared scenes and acoustic rendering.}
Eight bases in each of 100 exhibition rooms define speech, robot playback,
noise, initial geometry, and duration, up to 30~s.
Speech is drawn from a LibriSpeech \citep{panayotov2015librispeech} inventory of 300 speakers; screened non-speech recordings and synthetic noise
provide environmental interference.
Each base produces one static and two independently sampled dynamic
realizations (Dynamic A and B in Figure~\ref{fig:construction}c), preserving
content and initial layout.
SoundSpaces 2.0 \citep{chen2022soundspaces} renders FOA audio at every
receiver through room impulse responses (RIRs), which describe sound
propagation from each source to each receiver. Static RIRs or time-varying
RIRs are convolved with the source signals and mixed with environmental noise.
Each receiver's own playback is ideally removed; other robots' speech remains.
Dynamic paths are horizontal, with fixed height and robot yaw.
The paired realizations support comparison of shared content, but are
not single-factor motion interventions.

\paragraph{Queries and validation.}
Speech alignment supplies word and activity times; room sampling selects
the scene geometry, and noise screening excludes unsuitable recordings.
The scene truth branch in Figure~\ref{fig:construction}c supplies source
positions and speech-active intervals in each coordinate frame. These
labels and transcripts generate view-specific queries and reference answers. Source identities link the same speaker
across receivers; local coordinates are transformed for each egocentric view.
Automated checks cover audio reconstruction, noise calibration, motion
feasibility, coordinate consistency, and unique target selection after
prompt rounding. Dynamic egocentric transcription candidates with
insufficient relative motion are filtered before release. The manual
validation stage supports human listening and annotation review.
Appendix~\ref{app:human-review} describes the implemented interface and
procedure.

The resulting test set contains 800 static and 1,600 dynamic realizations,
4,761 receiver recordings, and approximately 40,000 questions across 13 tasks
(Appendix~\ref{app:data}).
Figure~\ref{fig:construction}d reports nominal signal-to-noise ratio (SNR)
over bases and room reverberation time (RT60, the time for sound energy
to decay by 60~dB) over rooms. Environmental noise occurs in 95\% of bases;
nominal SNR spans $-5$ to $25$~dB, while Clean means no added noise.
Received ratios vary by target and receiver. Scene composition counts
human--receiver population combinations; azimuth and distance plots count
3,182 initial human--receiver pairs once per base, showing spatial coverage
rather than the question distribution.
Appendix~\ref{app:data} gives construction parameters, distributions,
and validation scope.

\begin{table*}[!t]
\centering
\fontsize{8}{9.6}\selectfont
\setlength{\tabcolsep}{1.05pt}
\renewcommand{\arraystretch}{1.12}
\begin{tabular}{@{}lcccccccccc@{}}
\toprule
& & \multicolumn{5}{c}{Static} & \multicolumn{3}{c}{Dynamic} & \\
\cmidrule(lr){3-7}\cmidrule(lr){8-10}
System & Input & \taskDirectionEstimation{} & \taskDistanceEstimation{} & \taskTextDirection{} & \taskTextDistance{} & \taskEgoStaticASR{} & \taskDirectionTracking{} & \taskDistanceTracking{} & \taskEgoDynamicASR{} & Overall \\
Metric & & Acc$\uparrow$ & RelAcc$\uparrow$ & Acc$\uparrow$ & RelAcc$\uparrow$ & WER$\downarrow$ & Acc$\uparrow$ & RelAcc$\uparrow$ & WER$\downarrow$ & \\
\midrule
\multicolumn{11}{@{}l}{\textit{Single-channel encoding LALMs}} \\
Qwen3-Omni & FOA-W & 11.34 & 18.79 & 10.15 & 20.52 & 96.53 & 3.34 & 20.73 & 128.00 & 20.56 \\
Qwen3-Omni & FOA-sep & 13.07 & 19.61 & 8.71 & \textbf{21.68} & 93.55 & 4.07 & 19.52 & 123.70 & 21.41 \\
Qwen3.5-Omni-Plus & FOA-W & 18.13 & 21.34 & 9.37 & 20.77 & 80.74 & 6.56 & \textbf{22.24} & 78.00 & 24.75 \\
Qwen3.5-Omni-Plus & FOA-sep & 17.33 & \textbf{21.90} & 9.15 & 20.65 & 72.14 & 7.64 & 19.40 & \textbf{73.10} & 26.52 \\
\midrule
\multicolumn{11}{@{}l}{\textit{Multichannel encoding LALMs}} \\
BAT & Binaural & 0.00 & 0.00 & 0.00 & 0.00 & 99.89 & 0.00 & 0.00 & 99.82 & 0.07 \\
MiDasheng-Spatial & Binaural & 4.62 & 6.10 & 7.95 & 13.76 & 77.21 & 0.00 & 1.20 & 86.56 & 13.23 \\
Spatial-Omni-7B & FOA & 7.63 & 6.91 & 10.97 & 15.02 & 87.68 & 1.30 & 0.70 & 97.35 & 13.66 \\
Spatial-Omni-30B & FOA & 18.91 & 5.81 & 24.76 & 5.88 & 137.11 & 6.70 & 5.75 & 165.46 & 18.54 \\
JAEGER Neural & FOA & 13.62 & 7.23 & 7.54 & 3.33 & 101.62 & 0.00 & 3.35 & 101.20 & 2.80 \\
JAEGER Classical & FOA & 7.60 & 0.06 & 0.19 & 0.00 & 103.47 & 0.04 & 3.14 & 100.75 & 0.94 \\
Qwen3.8-Omni-Flash & FOA & 41.37 & 13.98 & \textbf{50.03} & 19.55 & 72.09 & \textbf{41.54} & 19.36 & 73.32 & 30.18 \\
\midrule
\multicolumn{11}{@{}l}{\textit{Agentic specialist tool-calling system}} \\
MESH-agent & FOA & \textbf{49.34} & 16.47 & 47.17 & 20.43 & \textbf{61.09} & 28.61 & 19.06 & 78.98 & \textbf{34.97} \\
\bottomrule
\end{tabular}
\vspace{-5pt}
\caption{Egocentric results (\%). Task codes follow Table~\ref{tab:task-reference}. Bold: column best. Overall uses Eq.~\ref{eq:overall}.}
\label{tab:egocentric-results}
\vspace{-5pt}
\end{table*}

\begin{table*}[!t]
\centering
\fontsize{8}{9.6}\selectfont
\setlength{\tabcolsep}{1.45pt}
\renewcommand{\arraystretch}{1.23}
\begin{minipage}[t]{0.64\textwidth}
\vspace{0pt}
\centering
\begin{tabular}{@{}lccccccc@{}}
\toprule
& & \multicolumn{3}{c}{Static} & \multicolumn{2}{c}{Dynamic} & \\
\cmidrule(lr){3-5}\cmidrule(lr){6-7}
System & Input & \taskSourceLocalization{} & \taskTextLocalization{} & \taskGlobalStaticASR{} & \taskSourceTracking{} & \taskGlobalDynamicASR{} & Overall \\
Metric & & Acc$\uparrow$ & Acc$\uparrow$ & WER$\downarrow$ & Acc$\uparrow$ & WER$\downarrow$ & \\
\midrule
\multicolumn{8}{@{}l}{\textit{Single-channel encoding LALMs}} \\
Qwen3-Omni & FOA-W & 1.19 & 0.31 & 104.00 & 6.48 & 95.10 & 13.77 \\
Qwen3-Omni & FOA-sep & 1.19 & 0.06 & 109.59 & 6.45 & 92.06 & 14.22 \\
Qwen3.5-Omni-Plus & FOA-W & 9.49 & 7.62 & \textbf{71.06} & 6.93 & 74.10 & 19.45 \\
Qwen3.5-Omni-Plus & FOA-sep & 10.99 & 11.37 & 72.65 & 6.41 & 68.91 & 21.86 \\
\midrule
\multicolumn{8}{@{}l}{\textit{Multichannel encoding LALMs}} \\
BAT & Binaural & 0.00 & 0.00 & 99.89 & 0.00 & 99.98 & 0.04 \\
MiDasheng-Spatial & Binaural & 7.00 & 7.56 & 83.36 & 0.19 & 81.18 & 15.36 \\
Spatial-Omni-7B & FOA & 11.62 & 6.43 & 96.39 & 0.28 & 89.87 & 12.55 \\
Spatial-Omni-30B & FOA & 2.87 & 3.31 & 150.43 & 2.66 & 152.76 & 12.72 \\
JAEGER Neural & FOA & 5.00 & 3.56 & 106.95 & 0.00 & 101.57 & 6.58 \\
JAEGER Classical & FOA & 0.00 & 0.06 & 102.87 & 0.00 & 101.38 & 1.21 \\
Qwen3.8-Omni-Flash & FOA & 3.75 & 4.43 & 73.98 & 4.19 & 73.68 & 15.90 \\
\midrule
\multicolumn{8}{@{}l}{\textit{Agentic specialist tool-calling system}} \\
MESH-agent & FOA & \textbf{32.04} & \textbf{31.79} & 78.54 & \textbf{18.39} & \textbf{59.15} & \textbf{31.01} \\
\bottomrule
\end{tabular}
\end{minipage}\hfill
\begin{minipage}[t]{0.34\textwidth}
\vspace{0pt}
\setlength{\tabcolsep}{3pt}
\renewcommand{\arraystretch}{1.06}
\centering
\setlength{\tabcolsep}{2pt}
\renewcommand{\arraystretch}{1.07}
\begin{tabularx}{\linewidth}{@{}lY@{}}
\toprule
Code & Full task name \\
\midrule
\multicolumn{2}{@{}l}{\textit{Egocentric: Static}} \\
\taskDirectionEstimation{} & Direction localization \\
\taskDistanceEstimation{} & Distance localization \\
\taskTextDirection{} & Transcript-conditioned direction localization \\
\taskTextDistance{} & Transcript-conditioned distance localization \\
\taskEgoStaticASR{} & Spatially cued transcription \\
\midrule
\multicolumn{2}{@{}l}{\textit{Egocentric: Dynamic}} \\
\taskDirectionTracking{} & Direction trajectory localization \\
\taskDistanceTracking{} & Distance trajectory localization \\
\taskEgoDynamicASR{} & Spatially cued transcription \\
\midrule
\multicolumn{2}{@{}l}{\textit{Allocentric: Static}} \\
\taskSourceLocalization{} & Source-set position localization \\
\taskTextLocalization{} & Transcript-conditioned position localization \\
\taskGlobalStaticASR{} & Region-cued transcription \\
\midrule
\multicolumn{2}{@{}l}{\textit{Allocentric: Dynamic}} \\
\taskSourceTracking{} & Transcript-conditioned trajectory localization \\
\taskGlobalDynamicASR{} & Trajectory-cued transcription \\
\bottomrule
\end{tabularx}

\end{minipage}
\vspace{-10pt}
\caption{Allocentric results (\%). Full task names on the right are grouped by coordinate view and motion setting. Bold: column best. Overall uses Eq.~\ref{eq:overall}.}
\label{tab:global-results}
\label{tab:task-reference}
\vspace{-10pt}
\end{table*}

\section{Experiments}
\label{sec:evaluation}
\label{sec:protocol}
\label{sec:experiments}

\subsection{Setup and Evaluation Protocol}
\label{sec:temporal}

\paragraph{Systems and inputs.}
We evaluate eleven audio-model configurations and MESH-agent on
19,117 static and 20,882 dynamic questions.
The audio models include Qwen3-Omni \citep{xu2025qwen3omni}
(FOA-W and FOA-sep), Spatial-Omni-7B and Spatial-Omni-30B
\citep{zhu2026spatialomni}, JAEGER Neural and Classical
\citep{liu2026jaeger}, BAT \citep{zheng2024bat},
MiDasheng-Spatial \citep{hu2026midashenglmspatial},
Qwen3.5-Omni-Plus (FOA-W and FOA-sep), and Qwen3.8-Omni-Flash.
MESH-agent uses a frozen Qwen3-14B \citep{yang2025qwen3technicalreport}
planner to coordinate specialist audio models and signal-processing tools.
FOA-W uses the omnidirectional channel, while FOA-sep supplies
WXYZ as separate audio clips.
Qwen3.8-Omni-Flash receives native four-channel FOA input.
All systems use only the recordings, metadata, and queries
permitted by each task.

\paragraph{Evaluation metrics.}
Spatial predictions are considered correct within $20^\circ$
direction error, 20\% relative distance error, or 2~m
world-position error.
Dynamic localization predicts absolute spatial values on a
0.5-s grid and is evaluated on speech-active frames, with
consistent source assignment across direction tracks.
Missing or invalid predictions count as failures.
Spatial transcription is evaluated by corpus word error rate (WER),
which reflects both target selection and speech recognition.
The 800 empty-region queries are evaluated separately by
empty-answer accuracy.

\paragraph{Overall score.}
We also report a question-level Overall score for each view,
giving equal weight to localization and transcription.
Let $\mathcal L_v$ and $\mathcal S_v$ denote their respective
question sets in view $v$, including both static and dynamic tasks:
\begin{equation}
\mathrm{Overall}_v=50\!\left(
\frac{\sum_{i\in\mathcal L_v} a_i}{|\mathcal L_v|}
+\frac{\sum_{i\in\mathcal S_v} s_i}{|\mathcal S_v|}
\right).
\label{eq:overall}
\end{equation}
Here $a_i$ is the fraction of correct targets or active frames
within question $i$, and $s_i=\max(0,1-\mathrm{WER}_i)$ for
nonempty transcripts, with WER expressed as a ratio.
For empty-region queries, $s_i$ is one for a correct empty
answer and zero otherwise.
Questions are equally weighted within each group, and the
resulting score ranges from 0 to 100.
We retain the original task metrics for detailed comparisons.

\paragraph{Analysis protocol.}
We analyze performance by human count, received SNR and SIR,
source distance, and motion.
The condition analysis covers six model configurations:
Qwen3-Omni (FOA-sep), Spatial-Omni-7B, Spatial-Omni-30B,
MiDasheng-Spatial, Qwen3.8-Omni-Flash, and
Qwen3.5-Omni-Plus (FOA-sep).
We further compare three models under different peer-metadata
and receiver-recording inputs on a fixed 100-room subset.
These comparisons preserve the same questions and targets,
and use room-clustered bootstrap intervals to assess
performance differences.

Further details on model configurations, MESH-agent,
scoring, and experimental analyses are provided in
Appendix ~\ref{app:models} and ~\ref{app:results}.

\subsection{Overall Performance}
\label{sec:main-results}

Tables~\ref{tab:egocentric-results} and~\ref{tab:global-results}
report results on eight egocentric and five allocentric tasks,
respectively, covering both static and dynamic settings.
Spatial tasks use threshold accuracy (Acc) or relative-distance
accuracy (RelAcc), while transcription tasks use word error rate (WER).

\vspace{-5pt}
\paragraph{Localization.}
Qwen3.8-Omni-Flash leads the direct audio models on egocentric
direction tasks, reaching 50.03\% on \taskTextDirection{} and
41.54\% on \taskDirectionTracking{}.
Yet its distance accuracies remain below 20\%, and its allocentric
localization accuracies below 5\%.
MESH-agent performs better on allocentric localization, reaching
32.04\% on \taskSourceLocalization{} and 18.39\% on
\taskSourceTracking{}.
These results show that accurate direction estimates do not
necessarily translate into accurate distances or world positions.

\paragraph{Transcription.}
The best WERs across the four transcription tasks remain
between 59.15\% and 73.98\%.
Qwen3.5-Omni-Plus leads several direct-model transcription tasks,
while MESH-agent achieves the lowest WER on
\taskEgoStaticASR{} (61.09\%) and
\taskGlobalDynamicASR{} (59.15\%).
MESH-agent reaches Overall scores of 34.97 and 31.01 in the
egocentric and allocentric views, respectively.
No direct model consistently leads across tasks, and better
speech recognition does not guarantee better spatial grounding.
Additional model comparisons are provided in
Appendix~\ref{app:new-models}.

\subsection{Ablations and Diagnostics}
\label{sec:peer-diagnostics}
This section highlights the key insights from our ablation studies and diagnostic analyses. Detailed evaluation results and additional examples are provided in Appendices ~\ref{app:diagnostics}, ~\ref{app:information-utilization} and ~\ref{app:case-traces}.

\paragraph{Peer information provides limited benefits.}
Removing peer metadata reduces Qwen3-Omni (FOA-sep)'s
\taskDirectionEstimation{} accuracy from 14.20\% to 9.34\%,
but improves some Spatial-Omni-7B results.
The effects vary across models and tasks, with no consistent
improvement from additional peer information.
Current systems do not reliably use shared robot metadata
to improve spatial understanding.

\paragraph{Distributed recordings offer selective gains.}
Using all receivers improves dynamic allocentric transcription
for some models.
For example, Qwen3-Omni (FOA-sep)'s WER decreases from
87.44\% to 82.54\%.
Spatial-Omni-7B shows a larger improvement, but most of its
gain comes from a few length-limited responses.
Spatial localization shows no consistent benefit from
additional recordings.
Thus, access to distributed acoustic observations does not
yet lead to reliable improvements across tasks.

\paragraph{Competing sources challenge spatial transcription.}
\label{sec:condition-diagnostics}
\begin{figure}[t]
\centering
\includegraphics[width=\columnwidth]{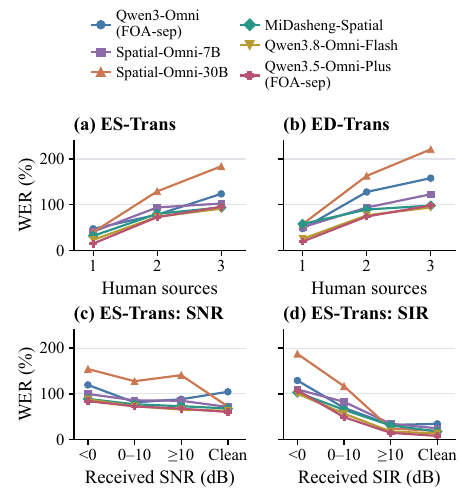}
\vspace{-10pt}
\caption{Egocentric transcription WER. Clean denotes no added noise (c) or no competing speech (d).}
\label{fig:condition-analysis}
\vspace{-10pt}
\end{figure}

Figure~\ref{fig:condition-analysis} shows that transcription
errors increase with the number of humans and stronger speech
interference.
More importantly, models often transcribe the wrong speaker.
For Qwen3.5-Omni-Plus (FOA-sep), 25.68\% of positive transcription
answers match another human's speech more closely than the target's.
Models also produce speech for empty target regions.
These failures show that recognizing speech and identifying
its physical source remain distinct challenges.

\paragraph{Distance and motion remain poorly grounded.}
\label{sec:motion-diagnostics}
Models frequently underestimate distant sources.
For targets at least 5~m away, Spatial-Omni-7B and
Spatial-Omni-30B predict a median distance of only 1.5~m,
compared with approximately 7.6~m in the ground truth.
Their predicted distance trajectories also remain nearly
constant in over 97\% of eligible cases, even when the true
distance changes.
As illustrated in Figure~\ref{fig:compact-cases}(b),
producing a trajectory does not necessarily mean tracking
the source's motion.

\section{Conclusion}
\label{sec:conclusion}

We introduce \bench{} to evaluate spatial speech understanding
through egocentric and allocentric views, covering 13 localization
and spatial transcription tasks under static and dynamic conditions.
Our results reveal persistent difficulties in associating speech
with the correct physical source, particularly in crowded scenes
and under motion.
More importantly, peer information and distributed recordings
provide only selective improvements, with no consistent gains
in spatial localization.
These findings highlight a gap between the availability of
spatial information and the ability of current systems to
use it effectively.
Future models should better integrate peer metadata, receiver
geometry, and acoustic observations across locations, while
maintaining source identity over time.
We release the dataset, construction pipeline, and evaluation
code to support the development of collaborative and distributed
spatial hearing systems.

\newpage
\section*{Limitations}

\paragraph{Data coverage.}
\bench{} uses simulated exhibition-room mixtures and English read
speech. Generalization to spontaneous conversations, other languages,
and physical recording environments remains untested.
Motion is restricted to horizontal paths with fixed height and robot
yaw, and long silent intervals are underrepresented.
RIR interpolation approximates time-varying propagation without
exact Doppler modeling, while recorded noise may retain its original
acoustics. Overlap with model pretraining data cannot be ruled out.
The preliminary listening audit covers only a small subset of scenes
and does not establish cross-receiver consistency.

\paragraph{Observation assumptions.}
Recordings are synchronized, receiver poses and playback logs are
known, and each receiver's own playback is ideally removed.
Source counts and target-selection cues are supplied where required.
Clock drift, uncertain calibration, missing recordings, and active
receiver control are not evaluated.
Offline evaluation does not establish real-time performance or
embedded deployment feasibility.

\paragraph{Experimental scope.}
Condition analyses cover selected models and do not independently
control acoustic factors.
Information ablations cover three models on a subset of scenes;
removing metadata also changes prompt length.
The reported confidence intervals do not account for multiple
comparisons.
MESH-agent was evaluated on a release with prior results, rather
than an untouched test set, and the contribution of agentic
planning has not been isolated on the full benchmark.
The observed errors do not uniquely identify their internal causes.

\section*{Ethical Considerations}

\paragraph{Potential risks.}
Spatial speech understanding can support assistive interfaces
and robots in shared environments, but may also enable
monitoring or speaker tracking without appropriate consent.
Incorrect speech--source associations can attribute an
utterance to the wrong person.
Benchmark performance does not establish suitability for
surveillance, personal identification, or consequential
decisions about individuals.
Deployment requires appropriate consent, privacy safeguards,
and validation in the intended environment.

\paragraph{Data and scope.}
The benchmark uses existing speech and noise recordings in
simulated acoustic scenes.
Scene-local target identifiers do not anonymize the underlying
voices, and data reuse remains subject to upstream licenses.
English read speech and simulated rooms do not establish
generalization across languages, demographic groups, or
physical environments.
Acoustic rendering and model inference require computational
resources; subsequent analyses reuse existing predictions
where possible to avoid unnecessary repeated inference.

\paragraph{Human review.}
A preliminary listening audit involved seven participants and
produced 57 completed review records covering 56 physical
realizations.
The audit included 36 pilot and 21 formal reviews.
Its limited coverage does not establish inter-rater agreement
or cross-receiver consistency.
The review procedure and observations are documented in
Appendix~\ref{app:human-review}.

\FloatBarrier
\bibliography{MESH_references_verified,manuscript/references_supplement}

\begin{thebibliography}{30}
\providecommand{\natexlab}[1]{#1}

\bibitem[{Biswas et~al.(2026)Biswas, Khan, and Islam}]{biswas2025owl}
Subrata Biswas, Mohammad Nur~Hossain Khan, and Bashima Islam. 2026.
\newblock \href {https://proceedings.iclr.cc/paper_files/paper/2026/hash/22f5d8e689d2a011cd8ead552ed59052-Abstract-Conference.html} {{OWL}: Geometry-aware spatial reasoning for audio large language models}.
\newblock In \emph{International Conference on Learning Representations}.

\bibitem[{Chen et~al.(2022)Chen, Schissler, Garg, Kobernik, Clegg, Calamia, Batra, Robinson, and Grauman}]{chen2022soundspaces}
Changan Chen, Carl Schissler, Sanchit Garg, Philip Kobernik, Alexander Clegg, Paul Calamia, Dhruv Batra, Philip Robinson, and Kristen Grauman. 2022.
\newblock \href {https://doi.org/10.52202/068431-0647} {{SoundSpaces 2.0: A Simulation Platform for Visual-Acoustic Learning}}.
\newblock In \emph{Advances in Neural Information Processing Systems}, volume~35, pages 8896--8911. Curran Associates, Inc.

\bibitem[{Chen et~al.(2026)Chen, Wang, Fan, Hao, Kong, Zhu, Hong, Chen, Su, Jian, Yang, Hu, Du, Luan, and Li}]{chen2026rmsaqa}
Peihao Chen, Qing Wang, Lichun Fan, Yufeng Hao, Zhifeng Kong, Mengyao Zhu, Hengyi Hong, Hang Chen, Hang Su, Yujie Jian, Chao-Han~Huck Yang, Shichao Hu, Jun Du, Jian Luan, and Ke~Li. 2026.
\newblock \href {https://arxiv.org/abs/2610.00935} {{RMS-AQA}: A two-stage spatial audio question answering benchmark for real-world domestic environments}.
\newblock \emph{arXiv preprint arXiv:2610.00935}.

\bibitem[{Cornell et~al.(2023)Cornell, Wiesner, Watanabe, Raj, Chang, Garcia, Maciejewski, Masuyama, Wang, Squartini, and Khudanpur}]{cornell2023chime7}
Samuele Cornell, Matthew Wiesner, Shinji Watanabe, Desh Raj, Xuankai Chang, Paola Garcia, Matthew Maciejewski, Yoshiki Masuyama, Zhong-Qiu Wang, Stefano Squartini, and Sanjeev Khudanpur. 2023.
\newblock \href {https://doi.org/10.21437/CHiME.2023-1} {{The CHiME-7 DASR Challenge: Distant Meeting Transcription with Multiple Devices in Diverse Scenarios}}.
\newblock In \emph{Proc. CHiME 2023}, pages 1--6.

\bibitem[{Evers et~al.(2020)Evers, L{\"o}llmann, Mellmann, Schmidt, Barfuss, Naylor, and Kellermann}]{evers2020locata}
Christine Evers, Heinrich~W. L{\"o}llmann, Heinrich Mellmann, Alexander Schmidt, Hendrik Barfuss, Patrick~A. Naylor, and Walter Kellermann. 2020.
\newblock \href {https://doi.org/10.1109/TASLP.2020.2990485} {The {LOCATA} challenge: Acoustic source localization and tracking}.
\newblock \emph{IEEE/ACM Transactions on Audio, Speech, and Language Processing}, 28:1620--1643.

\bibitem[{Field and Welsh(2007)}]{field2007clusterbootstrap}
C.~A. Field and A.~H. Welsh. 2007.
\newblock \href {https://doi.org/10.1111/j.1467-9868.2007.00593.x} {{Bootstrapping Clustered Data}}.
\newblock \emph{Journal of the Royal Statistical Society Series B: Statistical Methodology}, 69(3):369--390.

\bibitem[{Gardner and Martin(1995)}]{gardner1995kemar}
William~G. Gardner and Keith~D. Martin. 1995.
\newblock \href {https://doi.org/10.1121/1.412407} {{HRTF measurements of a KEMAR}}.
\newblock \emph{The Journal of the Acoustical Society of America}, 97(6):3907--3908.

\bibitem[{Guizzo et~al.(2022)Guizzo, Marinoni, Pennese, Ren, Zheng, Zhang, Masiero, Uncini, and Comminiello}]{guizzo2022l3das22}
Eric Guizzo, Christian Marinoni, Marco Pennese, Xinlei Ren, Xiguang Zheng, Chen Zhang, Bruno Masiero, Aurelio Uncini, and Danilo Comminiello. 2022.
\newblock \href {https://doi.org/10.1109/ICASSP43922.2022.9746872} {{L3DAS22} challenge: Learning {3D} audio sources in a real office environment}.
\newblock In \emph{IEEE International Conference on Acoustics, Speech and Signal Processing}, pages 9186--9190.

\bibitem[{Hu et~al.(2026)Hu, Su, Fan, Dinkel, Li, Dai, Zhang, Liu, Wang, Wu, Luan, Zou, and Qu}]{hu2026midashenglmspatial}
Jinbo Hu, Hang Su, Lichun Fan, Heinrich Dinkel, Gang Li, Zhanchen Dai, Yiru Zhang, Chang Liu, Peng Wang, Junnan Wu, Jian Luan, Cong Zou, and Heng Qu. 2026.
\newblock \href {https://arxiv.org/abs/2610.11156} {{MiDashengLM-Spatial: Unifying General Audio Understanding and Spatial Awareness}}.
\newblock \emph{arXiv preprint arXiv:2610.11156}.

\bibitem[{Kumar et~al.(2026)Kumar, Hersek, Dementyev, Pan, Chatterjee, Kumar, Duraiswami, Manocha, and Colaco}]{kumar2026sealm}
Sonal Kumar, Sinan Hersek, Artem Dementyev, Mengzhen Pan, Ishan Chatterjee, Anurag Kumar, Ramani Duraiswami, Dinesh Manocha, and Andrea Colaco. 2026.
\newblock \href {https://arxiv.org/abs/2610.05610} {{SEA-LM}: Egocentric spatial audio understanding for wearable microphone arrays}.
\newblock \emph{arXiv preprint arXiv:2610.05610}.

\bibitem[{Kwon et~al.(2023)Kwon, Li, Zhuang, Sheng, Zheng, Yu, Gonzalez, Zhang, and Stoica}]{kwon2023efficient}
Woosuk Kwon, Zhuohan Li, Siyuan Zhuang, Ying Sheng, Lianmin Zheng, Cody~Hao Yu, Joseph~E. Gonzalez, Hao Zhang, and Ion Stoica. 2023.
\newblock \href {https://doi.org/10.1145/3600006.3613165} {{Efficient Memory Management for Large Language Model Serving with {PagedAttention}}}.
\newblock In \emph{Proceedings of the ACM SIGOPS 29th Symposium on Operating Systems Principles}, pages 611--626. Association for Computing Machinery.

\bibitem[{Li et~al.(2026)Li, Xi, Jiang, Wang, Knill, Gales, Li, and Yu}]{li2026dae}
Haoyu Li, Yu~Xi, Yidi Jiang, Shuai Wang, Kate Knill, Mark Gales, Haizhou Li, and Kai Yu. 2026.
\newblock \href {https://doi.org/10.24963/ijcai.2026/644} {Detect, attend and extract: Keyword guided target speaker extraction}.
\newblock In \emph{Proceedings of the Thirty-Fifth International Joint Conference on Artificial Intelligence}, pages 5784--5792.

\bibitem[{Liu et~al.(2026{\natexlab{a}})Liu, Tang, Wang, Zhu, Chen, Shao, Wang, Ke, Jin, and Zhang}]{liu2026jaeger}
Zhan Liu, Changli Tang, Yuxin Wang, Zhiyuan Zhu, Youjun Chen, Yiwen Shao, Tianzi Wang, Lei Ke, Zengrui Jin, and Chao Zhang. 2026{\natexlab{a}}.
\newblock \href {https://proceedings.mlr.press/v306/liu26ea.html} {{JAEGER}: Joint {3D} audio-visual grounding and reasoning in simulated physical environments}.
\newblock In \emph{Proceedings of the 43rd International Conference on Machine Learning}, volume 306 of \emph{Proceedings of Machine Learning Research}, pages 78566--78578. PMLR.

\bibitem[{Liu et~al.(2026{\natexlab{b}})Liu, Niu, Xiao, Zheng, Yuan, Zang, Cao, Dong, Liang, Chen, Sun, Lin, and Wang}]{liu2026star}
Zihan Liu, Zhikang Niu, Qiuyang Xiao, Zhisheng Zheng, Ruoqi Yuan, Yuhang Zang, Yuhang Cao, Xiaoyi Dong, Jianze Liang, Xie Chen, Leilei Sun, Dahua Lin, and Jiaqi Wang. 2026{\natexlab{b}}.
\newblock \href {https://proceedings.iclr.cc/paper_files/paper/2026/hash/d9f8b5abc8e0926539ecbb492af7b2f1-Abstract-Conference.html} {{STAR-Bench}: Probing deep spatio-temporal reasoning as audio 4d intelligence}.
\newblock In \emph{International Conference on Learning Representations}, pages 134703--134731.

\bibitem[{Oh et~al.(2026)Oh, Shimada, Takida, Kim, Uesaka, Shibuya, Lee, Oh, and Mitsufuji}]{oh2026staudio}
Hyun-Bin Oh, Kazuki Shimada, Yuhta Takida, Sung-Bin Kim, Toshimitsu Uesaka, Takashi Shibuya, Kyeongyoon Lee, Tae-Hyun Oh, and Yuki Mitsufuji. 2026.
\newblock \href {https://arxiv.org/abs/2606.14141} {Spatio-temporal audio language modeling for dynamic sound sources}.
\newblock \emph{arXiv preprint arXiv:2606.14141}.

\bibitem[{Panayotov et~al.(2015)Panayotov, Chen, Povey, and Khudanpur}]{panayotov2015librispeech}
Vassil Panayotov, Guoguo Chen, Daniel Povey, and Sanjeev Khudanpur. 2015.
\newblock \href {https://doi.org/10.1109/icassp.2015.7178964} {{LibriSpeech: An ASR corpus based on public domain audio books}}.
\newblock In \emph{2015 IEEE International Conference on Acoustics, Speech and Signal Processing (ICASSP)}, pages 5206--5210.

\bibitem[{Polok et~al.(2026)Polok, Klement, Kocour, Han, Landini, Yusuf, Wiesner, Khudanpur, {\v{C}}ernock{\'y}, and Burget}]{polok2026dicow}
Alexander Polok, Dominik Klement, Martin Kocour, Jiangyu Han, Federico Landini, Bolaji Yusuf, Matthew Wiesner, Sanjeev Khudanpur, Jan {\v{C}}ernock{\'y}, and Luk{\'a}{\v{s}} Burget. 2026.
\newblock \href {https://doi.org/10.1016/j.csl.2025.101841} {{DiCoW: Diarization-conditioned Whisper for target speaker automatic speech recognition}}.
\newblock \emph{Computer Speech \& Language}, 95:101841.

\bibitem[{Sakshi et~al.(2025)Sakshi, Tyagi, Kumar, Seth, Selvakumar, Nieto, Duraiswami, Ghosh, and Manocha}]{sakshi2025mmau}
S~Sakshi, Utkarsh Tyagi, Sonal Kumar, Ashish Seth, Ramaneswaran Selvakumar, Oriol Nieto, Ramani Duraiswami, Sreyan Ghosh, and Dinesh Manocha. 2025.
\newblock \href {https://proceedings.iclr.cc/paper_files/paper/2025/file/d36f208919582785db965fe648b9fe59-Paper-Conference.pdf} {{MMAU}: A massive multi-task audio understanding and reasoning benchmark}.
\newblock In \emph{International Conference on Learning Representations}.

\bibitem[{Schmalenstroeer et~al.(2023)Schmalenstroeer, Gburrek, and Haeb-Umbach}]{schmalenstroeer2023libriwasn}
Joerg Schmalenstroeer, Tobias Gburrek, and Reinhold Haeb-Umbach. 2023.
\newblock \href {https://ris.uni-paderborn.de/record/48270} {{LibriWASN}: A data set for meeting separation, diarization, and recognition with asynchronous recording devices}.
\newblock In \emph{ITG Conference on Speech Communication}.

\bibitem[{Shi et~al.(2026)Shi, Wang, Guo, Wang, Zhang, Zhang, Guo, Hao, Xi, Yang, Xu, Zhou, and Lin}]{shi2026qwen3asr}
Xian Shi, Xiong Wang, Zhifang Guo, Yongqi Wang, Pei Zhang, Xinyu Zhang, Zishan Guo, Hongkun Hao, Yu~Xi, Baosong Yang, Jin Xu, Jingren Zhou, and Junyang Lin. 2026.
\newblock \href {https://arxiv.org/abs/2601.21337} {{Qwen3-ASR Technical Report}}.
\newblock \emph{arXiv preprint arXiv:2601.21337}.

\bibitem[{Shimada et~al.(2025)Shimada, Uchida, Koyama, Shibuya, Takahashi, Mitsufuji, and Kawahara}]{shimada2025ovseld}
Kazuki Shimada, Kengo Uchida, Yuichiro Koyama, Takashi Shibuya, Shusuke Takahashi, Yuki Mitsufuji, and Tatsuya Kawahara. 2025.
\newblock \href {https://doi.org/10.1109/taslpro.2025.3587455} {{Open-Vocabulary Sound Event Localization and Detection With Joint Learning of CLAP Embedding and Activity-Coupled Cartesian DOA Vector}}.
\newblock \emph{IEEE Transactions on Audio, Speech and Language Processing}, 33:2946--2960.

\bibitem[{Tang et~al.(2024)Tang, Yu, Sun, Chen, Tan, Li, Zhang, Lu, Ma, Wang, and Zhang}]{tang2024spatialaudio}
Changli Tang, Wenyi Yu, Guangzhi Sun, Xianzhao Chen, Tian Tan, Wei Li, Jun Zhang, Lu~Lu, Zejun Ma, Yuxuan Wang, and Chao Zhang. 2024.
\newblock \href {https://doi.org/10.21437/Interspeech.2024-2419} {Can large language models understand spatial audio?}
\newblock In \emph{Proceedings of Interspeech}, pages 4149--4153.

\bibitem[{Wang et~al.(2025)Wang, Zou, Lin, Sun, Liu, Zhang, Liu, Aw, and Chen}]{wang-etal-2025-audiobench}
Bin Wang, Xunlong Zou, Geyu Lin, Shuo Sun, Zhuohan Liu, Wenyu Zhang, Zhengyuan Liu, AiTi Aw, and Nancy~F. Chen. 2025.
\newblock \href {https://doi.org/10.18653/v1/2025.naacl-long.218} {{AudioBench}: A universal benchmark for audio large language models}.
\newblock In \emph{Proceedings of the 2025 Conference of the Nations of the Americas Chapter of the Association for Computational Linguistics: Human Language Technologies (Volume 1: Long Papers)}, pages 4297--4316. Association for Computational Linguistics.

\bibitem[{Wang et~al.(2026)Wang, Peng, Li, Wang, Tu, Xi, Sun, Yu, and Wang}]{wang2026audiomind}
Yucheng Wang, Jing Peng, Hanqi Li, Chenghao Wang, Wenming Tu, Yu~Xi, Zhaokai Sun, Kai Yu, and Shuai Wang. 2026.
\newblock \href {https://arxiv.org/abs/2605.28480} {{Audio-Mind: An Auditable Agentic Framework for Audio Understanding}}.
\newblock \emph{arXiv preprint arXiv:2605.28480}.

\bibitem[{Xu et~al.(2025)Xu, Guo, Hu, Chu, Wang, He, Wang, Shi, He, Zhu, Lv, Wang, Guo, Wang, Ma, Zhang, Zhang, Hao, Guo, Yang, Zhang, Ma, Wei, Bai, Chen, Liu, Wang, Yang, Liu, Ren, Zheng, Men, Zhou, Yu, Yang, Yu, Zhou, and Lin}]{xu2025qwen3omni}
Jin Xu, Zhifang Guo, Hangrui Hu, Yunfei Chu, Xiong Wang, Jinzheng He, Yuxuan Wang, Xian Shi, Ting He, Xinfa Zhu, Yuanjun Lv, Yongqi Wang, Dake Guo, He~Wang, Linhan Ma, Pei Zhang, Xinyu Zhang, Hongkun Hao, Zishan Guo, and 19 others. 2025.
\newblock \href {https://arxiv.org/abs/2509.17765} {{Qwen3-Omni Technical Report}}.
\newblock \emph{arXiv preprint arXiv:2509.17765}.

\bibitem[{Yang et~al.(2025{\natexlab{a}})Yang, Li, Yang, Zhang, Hui, Zheng, Yu, Gao, Huang, Lv, Zheng, Liu, Zhou, Huang, Hu, Ge, Wei, Lin, Tang, Yang, Tu, Zhang, Yang, Yang, Zhou, Zhou, Lin, Dang, Bao, Yang, Yu, Deng, Li, Xue, Li, Zhang, Wang, Zhu, Men, Gao, Liu, Luo, Li, Tang, Yin, Ren, Wang, Zhang, Ren, Fan, Su, Zhang, Zhang, Wan, Liu, Wang, Cui, Zhang, Zhou, and Qiu}]{yang2025qwen3technicalreport}
An~Yang, Anfeng Li, Baosong Yang, Beichen Zhang, Binyuan Hui, Bo~Zheng, Bowen Yu, Chang Gao, Chengen Huang, Chenxu Lv, Chujie Zheng, Dayiheng Liu, Fan Zhou, Fei Huang, Feng Hu, Hao Ge, Haoran Wei, Huan Lin, Jialong Tang, and 41 others. 2025{\natexlab{a}}.
\newblock \href {https://arxiv.org/abs/2505.09388} {Qwen3 technical report}.
\newblock \emph{Preprint}, arXiv:2505.09388.

\bibitem[{Yang et~al.(2025{\natexlab{b}})Yang, Li, Yang, Zhang, Hui, Zheng, Yu et~al.}]{qwen2025qwen3}
An~Yang, Anfeng Li, Baosong Yang, Beichen Zhang, Binyuan Hui, Bo~Zheng, Bowen Yu, et~al. 2025{\natexlab{b}}.
\newblock \href {https://arxiv.org/abs/2505.09388} {{Qwen3} technical report}.
\newblock \emph{arXiv preprint arXiv:2505.09388}.

\bibitem[{Yang et~al.(2024)Yang, Xu, Liu, Chu, Jiang, Zhou, Leng, Lv, Zhao, Zhou, and Zhou}]{yang-etal-2024-air}
Qian Yang, Jin Xu, Wenrui Liu, Yunfei Chu, Ziyue Jiang, Xiaohuan Zhou, Yichong Leng, Yuanjun Lv, Zhou Zhao, Chang Zhou, and Jingren Zhou. 2024.
\newblock \href {https://doi.org/10.18653/v1/2024.acl-long.109} {{AIR}-bench: Benchmarking large audio-language models via generative comprehension}.
\newblock In \emph{Proceedings of the 62nd Annual Meeting of the Association for Computational Linguistics (Volume 1: Long Papers)}, pages 1979--1998. Association for Computational Linguistics.

\bibitem[{Zheng et~al.(2024)Zheng, Peng, Ma, Chen, Choi, and Harwath}]{zheng2024bat}
Zhisheng Zheng, Puyuan Peng, Ziyang Ma, Xie Chen, Eunsol Choi, and David Harwath. 2024.
\newblock \href {https://proceedings.mlr.press/v235/zheng24i.html} {{BAT}: Learning to reason about spatial sounds with large language models}.
\newblock In \emph{Proceedings of the 41st International Conference on Machine Learning}, volume 235 of \emph{Proceedings of Machine Learning Research}, pages 61454--61469. PMLR.

\bibitem[{Zhu et~al.(2026)Zhu, Chen, Shao, Guo, Pan, Zhang, Wang, Liu, Zhang, Zeng, Cheng, Liu, Yang, Yves, Bo, and Zhao}]{zhu2026spatialomni}
Zhiyuan Zhu, Yixuan Chen, Yiwen Shao, Wenxiang Guo, Changhao Pan, Yu~Zhang, Yuxiang Wang, Wei Liu, Houhua Zhang, Chengkuan Zeng, Wenbo Cheng, Yunxi Liu, Rui Yang, Steve Yves, Liefeng Bo, and Zhou Zhao. 2026.
\newblock \href {https://arxiv.org/abs/2606.10738} {{Spatial-Omni}: Spatial audio understanding integration in multimodal {LLMs} via {FOA} encoding}.
\newblock \emph{arXiv preprint arXiv:2606.10738}.

\end{thebibliography}
\appendix
\section{Detailed Experimental Configurations}
\label{app:models}
Table~\ref{tab:model-inputs} summarizes the evaluated interfaces. This
appendix retains settings needed to interpret the comparisons; the release
archive contains full manifests, raw outputs, prompts, and detailed tables.
\subsection{LALM Inputs and Decoding}
\begin{table*}[t]
\centering\small
\setlength{\tabcolsep}{4pt}
\begin{tabularx}{\textwidth}{@{}l l Y@{}}
\toprule
System & Input & Recorded configuration \\
\midrule
Qwen3-Omni & FOA-W, FOA-sep & Qwen3-Omni-30B-A3B-Instruct; BF16, greedy; 1,024 or 8,192 output tokens \\
Qwen3.5-Omni-Plus & FOA-W, FOA-sep & Closed-source API; W only or four separate WXYZ inputs \\
Spatial-Omni-7B & FOA & Spatial-Omni-7B and SO encoder; labeled blocks up to 20 s; BF16, greedy \\
Spatial-Omni-30B & FOA & Native spatial and semantic frontends; input and short-decoder reference checks \\
JAEGER Neural & FOA & Neural IV, audio-only checkpoint 2; BF16, greedy \\
JAEGER Classical & FOA & Classical IV, audio-only checkpoint 2; native input-feature audit \\
BAT & Binaural & KEMAR conversion; 32-kHz, labeled blocks up to 10 s; BF16, greedy \\
MiDasheng-Spatial & Binaural & KEMAR conversion; full clips; FP32, greedy, repetition penalty 1.05 \\
Qwen3.8-Omni-Flash & FOA & Closed-source API; native four-channel FOA input \\
MESH-agent & FOA & Qwen3-14B planner; eight steps, four logical acoustic calls, 14,000 input tokens \\

\bottomrule
\end{tabularx}
\caption{Recorded input and inference configurations.}
\label{tab:model-inputs}
\end{table*}

Qwen3.8-Omni-Flash and Qwen3.5-Omni-Plus are closed-source API models.
The other audio models use open-source releases.
FOA-W supplies the omnidirectional channel; FOA-sep encodes WXYZ as
four labeled mono inputs. This does not establish native joint FOA processing.
Spatial-Omni-7B and Spatial-Omni-30B retain their spatial and semantic frontends.
BAT and MiDasheng-Spatial receive KEMAR binaural decodings, without independent
channel normalization. KEMAR measurements are described by
\citet{gardner1995kemar}. Receiver labels and original clip timing are retained.

The frozen Qwen3-Omni, Spatial-Omni-7B, JAEGER Neural, and BAT workers use greedy
decoding, temperature zero, seed zero, and BF16 language inference.
The output budget is 1,024 tokens for static and transcription outputs
and 8,192 for trajectories, subject to model context limits.
Spatial-Omni-7B uses labeled temporal blocks of at most 20~s; BAT uses 32-kHz
blocks of at most 10~s. These blocks are not additional microphones.
MiDasheng-Spatial uses \texttt{mispeech/midashenglm-spatial} revision
\texttt{160ae51b51be}, official code \texttt{e12bc10b1279}, FP32 vLLM
0.22 \citep{kwon2023efficient}, greedy decoding, and repetition penalty 1.05 with the same token budgets.
Its full clips retain the 185-sample filter tail without extending the task grid.
KEMAR decoding differs from its ARI training distribution, and multi-receiver
blocks extend its native evaluation setting; these differences limit attribution.

\paragraph{JAEGER checkpoints.}
JAEGER Neural and JAEGER Classical use the authors' released
\href{https://huggingface.co/tsinghua-ee/JAEGER/tree/main/checkpoints/neural_iv/audio_only_2}{\texttt{neural\_iv/audio\_only\_2}}
and \href{https://huggingface.co/tsinghua-ee/JAEGER/tree/main/checkpoints/classical_iv/audio_only_2}{\texttt{classical\_iv/audio\_only\_2}}
checkpoints, respectively, from the
\href{https://huggingface.co/tsinghua-ee/JAEGER}{JAEGER repository}.
OWL is excluded because its release does not establish a reproducible
match to the final CoT-trained model; a historical ST-AudioLM run is excluded
because its FOA basis conversion was incorrect.

\subsection{Input and Output Audits}
\label{app:input-audits}
Native reference checks match six Spatial-Omni-30B inputs, eight MiDasheng-Spatial inputs,
and six JAEGER Classical feature tensors. Short decoder checks validate
those examples, not all full-length answers.
\paragraph{Additional archived inference settings.}
The archived Qwen3-Omni worker configurations use tensor parallelism of
one, a maximum context length of 32,768 tokens, and a batched-token budget
of 8,192. Prefix caching is enabled, with at most 128 sequences and a GPU
memory utilization setting of 0.96. The multimodal input limits allow up
to three audio blocks for FOA-W and twelve for FOA-sep, each containing
at most 480,000 samples. These are interface limits: the loader supplies
only the recordings permitted by the current query. The Spatial-Omni-7B records
identify the Spatial-Omni-7B and SO-Encoder weight files by SHA-256, separately from
the Qwen2.5-Omni-7B text decoder.

\subsection{Agentic Specialist Tool-Calling System}
\label{app:agent-configuration}
MESH-agent denotes frozen Agent V2, with Qwen3-14B \citep{qwen2025qwen3} revision
\texttt{40c069824f42} as its planner. Capability cards and request-local
memory guide structured calls; the planner selects observations and tool
transcripts rather than inventing numerical observations or rewriting transcripts.
The tools include S5, OV-Seld \citep{shimada2025ovseld}, the Jiang distance estimator,
DiCoW \citep{polok2026dicow}, Spatial-Omni distance alternatives,
Qwen3-ASR-1.7B \citep{shi2026qwen3asr},
direction-conditioned FOA MVDR/cardioid filtering, and coordinate transforms.
Source slots and activity are predicted, not supplied by reference enrollment.
Each question permits eight planner steps, four logical acoustic calls,
and 14,000 input tokens per call. A logical call may invoke several models
or receivers and is not a neural forward-pass count.
The policy was frozen on development data; cached observations and recorded
execution passed deterministic replay checks. Earlier evaluation on this
release makes this a frozen regression, not an untouched-test evaluation.
Cached replay also does not measure complete online latency.

\paragraph{Capability cards and execution contract.}
The frozen system uses three task-specific card sets: the static
\texttt{v13} cards, the temporal
\texttt{exhibition-temporal-v1-public-protocol} cards, and the
\texttt{exhibition-dev-v2-manual} cards for ES-CDist, ES-Trans,
ED-Trans, and AD-Trans. Each card specifies its purpose, applicable
tasks, prerequisites, argument schema, returned evidence, and limitations.
The V2 extension exposes source observation, source-conditioned
recognition, spatially conditioned transcription, mixture observation,
source-range refinement, range estimation, candidate inspection, answer
selection, and failure reporting. Available tools and candidate references
are restricted by the task and the current observation state.
Arguments are checked against JSON schemas before execution; invalid
calls produce feedback rather than silently repaired observations.
Candidate transcripts remain evidence rather than instructions, and
memory is confined to the current question. Terminal selection returns
an observed candidate value without rewriting its transcript.

\paragraph{Planner runtime.}
Qwen3-14B runs in BF16 with vLLM 0.10.2 and tensor parallelism of two,
without quantization or CPU offloading. Generation uses temperature zero,
seed 290930, and at most 192 output tokens per planner call, with thinking
disabled. This output budget applies to the structured tool call; a
selected transcript is subsequently returned by the executor.
The server context limit is 16,384 tokens, while each planner input is
limited to 14,000 tokens. Calls use the native Hermes tool format with
JSON-schema-constrained decoding. The decoder omits the unsupported
\texttt{uniqueItems} grammar constraint, while the full prompt schema
and executor retain the uniqueness requirement.

\paragraph{Acoustic execution and caching.}
The additional Qwen3-ASR-1.7B backend uses BF16 inference with greedy
decoding, English as the requested language, an empty text context, and
a maximum output length of 512 tokens. It processes complete predicted
source waveforms or direction-conditioned recordings; reference transcripts
are not supplied as recognition prompts. The full evaluation used
40 RTX 4090 GPUs, running the additional ASR in 40 shards before launching
20 two-GPU planner replicas. Existing specialist observations were reused
from frozen caches. Source- and cue-conditioned ASR outputs were sealed
before fresh planner inference, and scoring followed prediction freezing
and deterministic replay checks. This execution arrangement evaluates
answer quality using cached perception and does not establish live
audio-to-answer latency.

\section{Detailed Results and Diagnostic Analyses}
\label{app:results}
This appendix gives the metric definitions and diagnostic evidence behind
the main-text findings on source competition, spatial errors, and selective
use of additional information. 

\subsection{Metrics and Reporting Scope}
\label{app:metrics}\label{app:overall}\label{app:exhibition-results}
For unit directions $\hat u,u$, distances $\hat r,r>0$, and positions
$\hat x,x$, the respective errors are
\begin{align}
e_\theta&=180\arccos(\operatorname{clip}(\hat u^\top u,-1,1))/\pi,\\
e_r&=|\hat r-r|/r,\qquad e_p=\|\hat x-x\|_2.
\end{align}
Thresholds follow Section~\ref{sec:temporal}.
Source sets use one-to-one matching; ED-Dir uses one assignment throughout
the clip. Direction and world-position scores pool reference positions,
using speech-active frames for tracks. ED-Dist first averages active-frame
success within each question, then weights questions equally.
Missing predictions fail. Coverage is the fraction of references with an
evaluable prediction, using the same aggregation. Missing directions receive
$180^\circ$ diagnostic error; distance and position errors condition on
valid predictions. Thus, low conditional error alone need not imply success.

Corpus WER is $100\sum_i(S_i+D_i+I_i)/\sum_i N_i$, using the person
selected by the query. Failed positive answers incur complete deletion;
repetitions and insertions remain. The 800 empty-region queries have no WER
denominator and instead require a valid empty answer; refusal is not absence.
Equation~\ref{eq:overall} averages localization accuracy within each question
and clips transcription scores \emph{before} averaging,
$s_i=\max(0,1-(S_i+D_i+I_i)/N_i)$, or the correct-empty indicator.
This differs from clipping corpus WER or averaging task means.
The weights were adopted after results were available and fixed across
models. Computations retain the frozen record-level denominators behind
the published scale of 40,000 questions.

\paragraph{Empty targets and Agent components.}
Correct absence affects rankings: excluding empty regions would change
Allocentric Overall from 14.22 to 16.10 for Qwen3-Omni (FOA-sep) and
from 15.36 to 14.19 for MiDasheng-Spatial.
MESH-agent's independently checked question-level archive uses the same
protocol; its localization and semantic components are shown in
Table~\ref{tab:case-equal-overall}. Its empty-answer accuracy is 22.50\%.
\begin{table}[t]
\centering
\footnotesize
\setlength{\tabcolsep}{4pt}
\renewcommand{\arraystretch}{1.12}
\begin{tabular}{@{}lrrr@{}}
\toprule
View & Localization & Semantic & Overall \\
\midrule
Egocentric & 27.85 & 42.10 & 34.97 \\
Allocentric & 23.67 & 38.35 & 31.01 \\
\bottomrule
\end{tabular}
\caption{MESH-agent Overall components (0--100; higher is better), including empty targets.}
\label{tab:case-equal-overall}
\end{table}

\paragraph{Qwen3.5-Omni-Plus components and uncertainty.}
The main tables report both interfaces' Overall scores.
Correct-empty accuracy is 6.00\% (FOA-W) and 12.13\% (FOA-sep).
The FOA-sep-minus-FOA-W Overall differences have room-bootstrap intervals
$[1.40,2.15]$ and $[1.95,2.91]$ points, respectively.
Against Flash, FOA-sep is lower Egocentrically ($-3.66$; $[-4.35,-3.01]$)
and higher Allocentrically ($+5.96$; $[5.33,6.59]$).
These intervals describe scene sampling for the evaluated systems.

\subsection{Condition Strata and Failure Distributions}
\label{app:diagnostics}\label{app:new-models}
\paragraph{Condition definitions.}
The human count $H$ is the number present, not simultaneous speakers;
scenes with $H=1$ can still contain robot playback. SNR compares target W-channel power
with environmental noise; SIR compares it with summed competing speech,
including waveform cross-terms. Powers use emitted target-active samples,
after ideal self-playback cancellation, before conversion to dB.
Finite bins are $<0$, $[0,10)$, and $\geq10$~dB. In
Figure~\ref{fig:condition-analysis}, Clean means no environmental noise
in (c), and no competing speech in (d); the connecting lines show ordered
conditions, not a continuous numerical interval to Clean.
The no-noise group has 134 static questions from 40 bases in 34 rooms;
the SIR $\geq10$~dB group has 72 questions from 43 rooms. Different group
compositions preclude a universal monotonicity or causal interpretation.

Selected differences use 2,000 room-cluster bootstrap replicates
\citep{field2007clusterbootstrap},
recomputing corpus WER from summed edits and reference words, with paired
room multiplicities and seed 20261012. The 95\% percentile intervals are
unadjusted for multiple comparisons and reflect scene sampling, not reruns.
All twelve three-human-minus-one-human intervals in the main figure are positive. The SIR contrasts
are also consistently adverse, while SNR and noise-type rankings differ by
model. Full contrasts and denominators are retained in the archive.

\begin{figure*}[t]
\centering
\includegraphics[width=\textwidth]{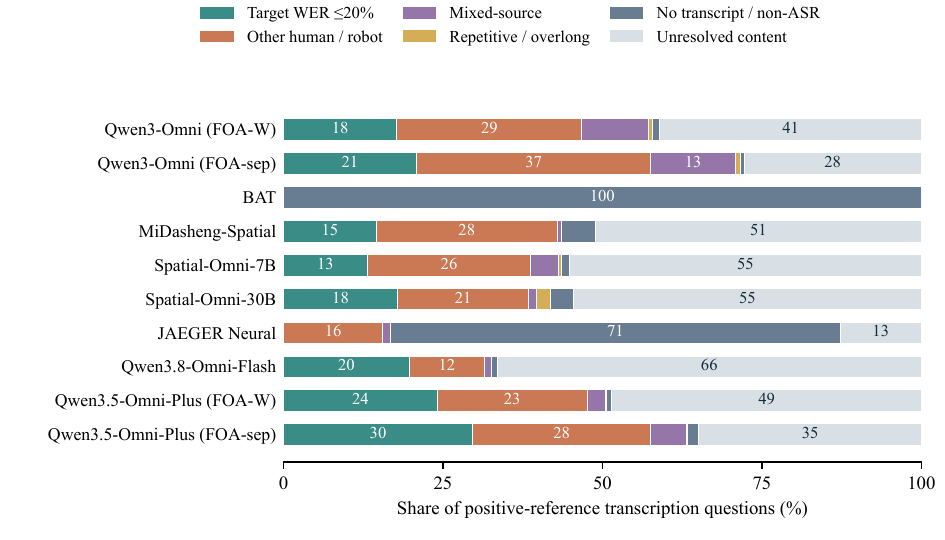}
\caption{Transcription outcome distribution over 12,925 positive questions per configuration. Mutually exclusive priority labels are grouped for display; the text defines their limits.}
\label{fig:asr-failure-distribution}
\end{figure*}

\paragraph{Transcription outcomes.}
Figure~\ref{fig:asr-failure-distribution} uses 12,925 positive-reference
questions per configuration. Priority labels are mutually exclusive:
no usable transcript; non-transcription answer; repetition; other-source
match; mixed-source evidence; exact or near target match; other overlong
output; unresolved content error. They are diagnostic categories, not a
causal decomposition. Classical and MESH-agent lack comparable full profiles.
An other-source match requires at least five words, candidate WER at most
35\%, target WER at least 65\%, and a gap of at least 30 points.
Human reference transcripts are hidden, but robot playback
texts are public, so robot-text matches do not identify listening versus copying.
Repetition requires at least 100 words, ten occurrences of one trigram,
and a distinct-trigram ratio below 0.35. Spatial-Omni-30B's 275 flagged answers
(2.13\%) contribute 49.21\% of its edits. This differs from reaching a token limit.

\paragraph{Distance and motion.}
Far-target medians condition on a valid scalar and true distance at least
5~m. Nearly constant ED-Dist outputs have predicted span below 0.01~m,
at least three readable active distances, and reference span above 0.5~m.
This excludes missing or short outputs: Spatial-Omni-7B and Spatial-Omni-30B have 1,576 and
4,176 eligible questions, of which 98.54\% and 97.49\% are constant.
Flash has 670 constant outputs among 4,537 eligible questions (14.77\%);
Plus FOA-sep has 1,125 among 4,035 (27.88\%). Their ED-Dist accuracies
remain 19.36\% and 19.40\%, respectively.
Matched static--dynamic comparisons average dynamic variants within each
base--human--receiver group, then weight groups equally. Flash's 2,999
matched groups show a 4.76-point increase in mean question WER
(95\% room interval 3.57--5.93); changing cues prevent a motion-only claim.

\subsection{Controlled Information Comparisons}
\label{app:information-utilization}\label{app:peer-ablation}
The paired study selects one base per room by fixed hash before scoring:
5,090 questions across 100 rooms, with 28,217 requests per model across
variants. All 84,651 requests are complete. Qwen3-Omni (FOA-sep), Spatial-Omni-7B,
and Spatial-Omni-30B retain fixed adapters, greedy decoding, and token budgets.
Each egocentric variant has 3,714 questions. E0 keeps full inputs;
E1 removes peer poses; E2 playback text; E3 playback times; E4 all playback;
E5 peer poses and playback. Deletions affect both structured inputs and
prompts, retaining recording, calibration, query, and target. Shorter prompts
are a confound; no length-matched placebo is included.
Allocentric variants retain all public metadata while changing only recordings.
The control is fixed R01, selected before scoring; fusion contrasts require
at least two receivers. Singleton, pair, and full sets retain receiver bindings.
Primary contrasts are E0 versus E5 and all receivers versus R01, with
2,000 paired room-bootstrap replicates and unadjusted 95\% intervals.
These subset scores do not enter full-test Overall.
\begin{figure*}[t]
\centering
\includegraphics[width=\textwidth]{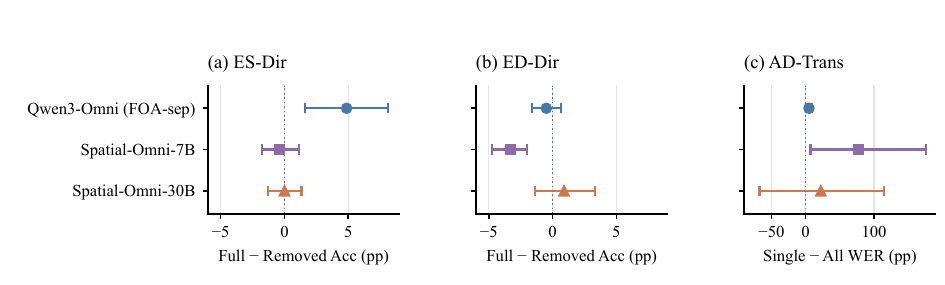}
\caption{Selected paired information effects with 95\% room-bootstrap intervals. Positive values favor keeping metadata (a,b) or adding recordings (c).}
\label{fig:information-contrasts}
\end{figure*}

\begin{table*}[t]
\centering
\footnotesize
\begin{tabularx}{\linewidth}{@{}Ycrrrr@{}}
\toprule
& & \multicolumn{2}{c}{Tuning DEV} & \multicolumn{2}{c}{Development holdout} \\
\cmidrule(lr){3-4}\cmidrule(lr){5-6}
Task & Metric (\%) & MESH-agent & Fixed MVDR & MESH-agent & Fixed MVDR \\
\midrule
Static egocentric \taskTextDistance{} & RelAcc@20\%$\uparrow$ & 21.92 & 23.85 & 19.19 & 20.35 \\
Static egocentric \taskEgoStaticASR{} & WER$\downarrow$ & 98.29 & 87.35 & 116.16 & 91.69 \\
Dynamic egocentric \taskEgoDynamicASR{} & WER$\downarrow$ & 106.81 & 86.29 & 94.77 & 92.03 \\
Dynamic allocentric \taskGlobalDynamicASR{} & WER$\downarrow$ & 81.79 & 70.65 & 82.64 & 69.90 \\
\bottomrule
\end{tabularx}
\caption{Separate development comparison using the same specialist observations. The fixed MVDR cascade makes no planner calls.}
\label{tab:agent-dev}
\end{table*}

The main text reports both benefits and harms from metadata deletion.
For the 270 paired AD-Trans questions, Qwen3-Omni improves on 65,
worsens on 38, and ties on 167 with all receivers; neither input reaches
the token limit. Spatial-Omni-7B improves on 93, worsens on 71, and ties on 106,
but six length-limited answers account for 71.44 of its 77.68-point WER
reduction. These decompositions retain every answer and the full denominator.
The separate earlier Spatial-Omni-30B study uses 12,891 questions per context;
all 16 contrast intervals contain zero. Its different population and newly
measured baseline are not mixed into the paired subset above.

\paragraph{Geometric opportunity versus acoustic evidence.}
On 135 AS-CPos questions, intersecting true source bearings perturbed by
5-degree RMS noise improves Acc@2m from 47.56\% for a fixed-R01 ray
with a 5-m depth prior to 89.07\% with all rays (difference interval
32.82--50.25 points). This is an oracle-bearing diagnostic, not an acoustic
baseline. A mean FOA-intensity frontend gated by true target-exclusive
activity improves only from 8.89\% to 11.85\% on the same full population
(interval $-2.96$--9.52), counting missing rays as failures. Only 70 questions
have sufficient exclusive activity at every receiver. The contrast identifies
geometric potential without isolating acoustic estimation, source association,
and fusion as independent causes of model failure.

\subsection{Separate Agent Development Comparison}
\label{app:agent}
A later round uses 1,520 tuning questions in 12 rooms and 992 holdout
questions in eight rooms. Three candidate changes to memory, verification,
and capability cards fail the selection criteria, so V2 remains frozen.
The fixed MVDR cascade outperforms it on all four measured holdout tasks
(Table~\ref{tab:agent-dev}).
The holdout retains 44 interrupted requests as failures (948 complete returns),
so it does not establish uninterrupted generalization. This development
population is separate from the main evaluation. The four displayed tasks
use 1,120 tuning and 720 holdout questions; the remaining 400 and 272
questions provide regression checks.

\section{Implementation, Prompts, and Human Review}
\label{app:implementation}
\subsection{Task Contracts and Coordinates}
\label{app:contracts}
\begin{table*}[t]
\centering
\small
\setlength{\tabcolsep}{4pt}
\renewcommand{\arraystretch}{1.16}
\begin{tabularx}{\linewidth}{@{}L{0.12\linewidth}L{0.105\linewidth}L{0.225\linewidth}Y L{0.20\linewidth}@{}}
\toprule
\textbf{View} & \textbf{Motion Setting} & \textbf{Task} & \textbf{Input} & \textbf{Required Output} \\
\midrule
\textbf{Egocentric} & Static & \taskDirectionEstimation{} & Human and peer-playback count & Direction set \\
\addlinespace[2pt]
& & \taskDistanceEstimation{} & Unique human azimuth sector & Target distance \\
\addlinespace[2pt]
& & \taskTextDirection{} & Target transcript & Target direction \\
\addlinespace[2pt]
& & \taskTextDistance{} & Target transcript & Target distance \\
\addlinespace[2pt]
& & \taskEgoStaticASR{} & Target direction and distance & Transcript \\
\cmidrule(l){2-5}
& Dynamic & \taskDirectionTracking{} & Human count $H$
& $H$ direction tracks \\
\addlinespace[2pt]
& & \taskDistanceTracking{} & Initial speech interval and azimuth sector & Target distance track \\
\addlinespace[2pt]
& & \taskEgoDynamicASR{} & Two local spatial anchors & Transcript \\
\midrule
\textbf{Allocentric} & Static & \taskSourceLocalization{} & Human count $H$
& Set of $H$ world positions \\
\addlinespace[2pt]
& & \taskTextLocalization{} & Target transcript & Target position \\
\addlinespace[2pt]
& & \taskGlobalStaticASR{} & World-coordinate region & Transcript \\
\cmidrule(l){2-5}
& Dynamic & \taskSourceTracking{} & Target transcript & Target position track \\
\addlinespace[2pt]
& & \taskGlobalDynamicASR{} & Target world trajectory on a 0.5-s grid & Transcript \\
\bottomrule
\end{tabularx}
\caption{Tasks by view and motion setting. Input lists cues added to the recordings and public context in Eqs.~\ref{eq:egocentric-inference}--\ref{eq:global-inference}. Transcript means all target speech in the clip.}
\label{tab:task-overview}
\end{table*}

Table~\ref{tab:task-overview} gives all task inputs and outputs using the
main-text codes. Dataset identifiers remain unchanged: \texttt{edge} and
\texttt{cloud} mean egocentric and allocentric. A FOA bundle is one microphone
location; temporal chunks are not additional receivers.
Directions use azimuth zero in front, positive left, and elevation positive
up. World $+y$ is up; zero robot yaw faces $-z$ and positive yaw turns
toward $+x$. Receiver poses transform world displacements into local
front--left--up coordinates. Angles are degrees; distance and position are meters.
Spatial tracks contain every requested 0.5-s timestamp and a speech-activity
flag; active frames carry spatial values, inactive frames omit them.
ED-Dir returns $H$ tracks with stable anonymous IDs, and AS-Pos
returns an unordered set of $H$ world positions. AD-Trans receives a full world
trajectory including silence, without activity labels, and returns the full transcript.

\subsection{Construction and Validation}
\label{app:data}
The benchmark scale is 40,000 questions drawn from a 300-speaker
LibriSpeech inventory, with 800 bases in 100 rooms, 800 static and
1,600 independently sampled dynamic realizations, and 4,761 FOA recordings.
Within a base, realizations share speech crops, playback, noise, initial
layout, and duration (mean 11.71~s); different bases use distinct recordings.
R01 is silent; other receivers may play speech, with ideal self-playback removal.
Signals are 16-kHz float32 ACN/SN3D FOA $[W,Y,Z,X]$.
Of 800 bases, 760 add noise (722 recorded and 38 synthetic); 40 are clean.
Noise gain is calibrated at static H01--R01 and retained across realizations,
so received target SNR and SIR vary. Common peak scaling preserves relative gains.
Prescribed horizontal minimum-jerk paths span 1.5--5~m at candidate peak
speeds 0.4--0.8~m/s, with fixed height and robot yaw. RIR knots every 0.2~s
and two-time interpolation retain propagation delays and reverberation tails.
Automated checks cover audio, labels, timestamps, prompt bindings, target
correspondence, and unique selection after coordinate rounding.
Filtering removes 1,424 ambiguous dynamic transcription candidates while
retaining their recordings. Oracle and missing-output tests check scoring.
These checks do not establish listening agreement or real-device generalization.

\subsection{Prompts and Answer Extraction}
\label{app:prompts}
The loader supplies the current query, permitted recordings and robot metadata;
other questions, scene images, and evaluator-only human labels are hidden.
The task table specifies the variable cues inserted into each prompt.
The following example preserves the actual wording of the frozen ES-Trans
query; full prompts for all thirteen tasks and receiver bindings are archived.
\begin{figure}[t]
\centering
\fcolorbox{blue!35!black}{blue!3}{\begin{minipage}{0.92\linewidth}
\small\raggedright
\textbf{ES-Trans: actual query}\par\smallskip
Select the same HUMAN speaker satisfying ALL of the following spatial conditions.
Throughout this static recording: direction [azimuth,elevation] [-164,11] degrees, spherical angular tolerance 10 degrees; distance 2.2 m, tolerance +/-0.5 m.
Exactly one human satisfies this query. Ignore other people, robot playback and environmental noise.
Transcribe ALL of that person's speech in the ENTIRE input.
Return only the English transcript; no timestamps, speaker IDs, JSON fields or explanation.
\end{minipage}}
\caption{A frozen task query, after shared public context.}
\label{fig:compact-prompt}
\end{figure}

Gold-blind extraction removes answer wrappers without correcting predicted
words or numbers. WER uses the requested target, never the best-matching
scene speaker. The frozen English tokenizer lowercases and uses its original
word-pattern rule; numbers are not expanded. Repetition is retained.
Spatial repair recovers existing fields only. One inherited manual extraction
remains hash-checked in the archive and is not an exhaustive listening review.
MESH-agent submits its selected tool transcript without rewriting.

\subsection{Human Review Interface and Procedure}
\label{app:human-review}
\begin{figure*}[tp]
\centering
\includegraphics[width=\textwidth]{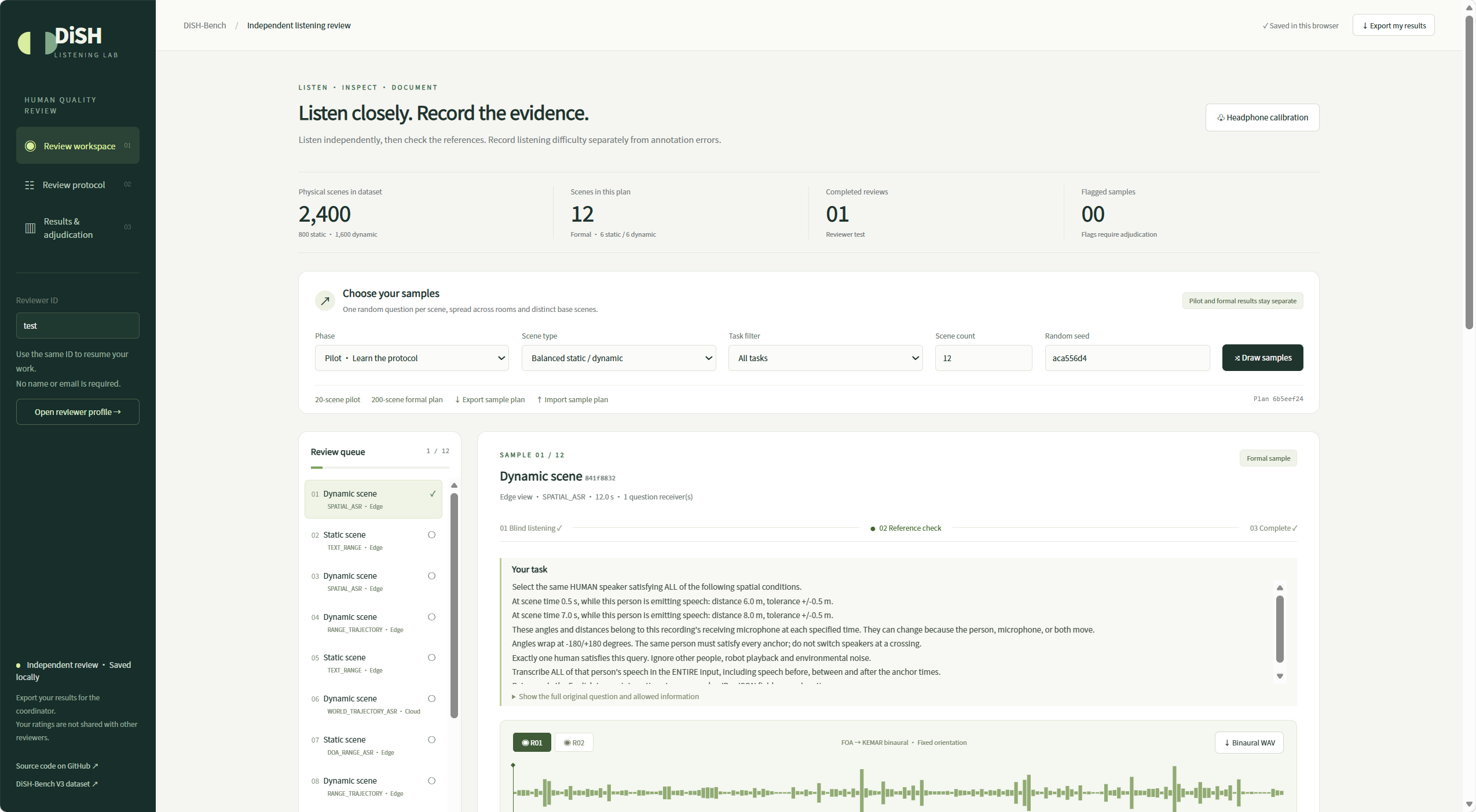}
\par\smallskip
{\small (a) Sampling and listening workspace.}
\par\medskip
\includegraphics[width=\textwidth]{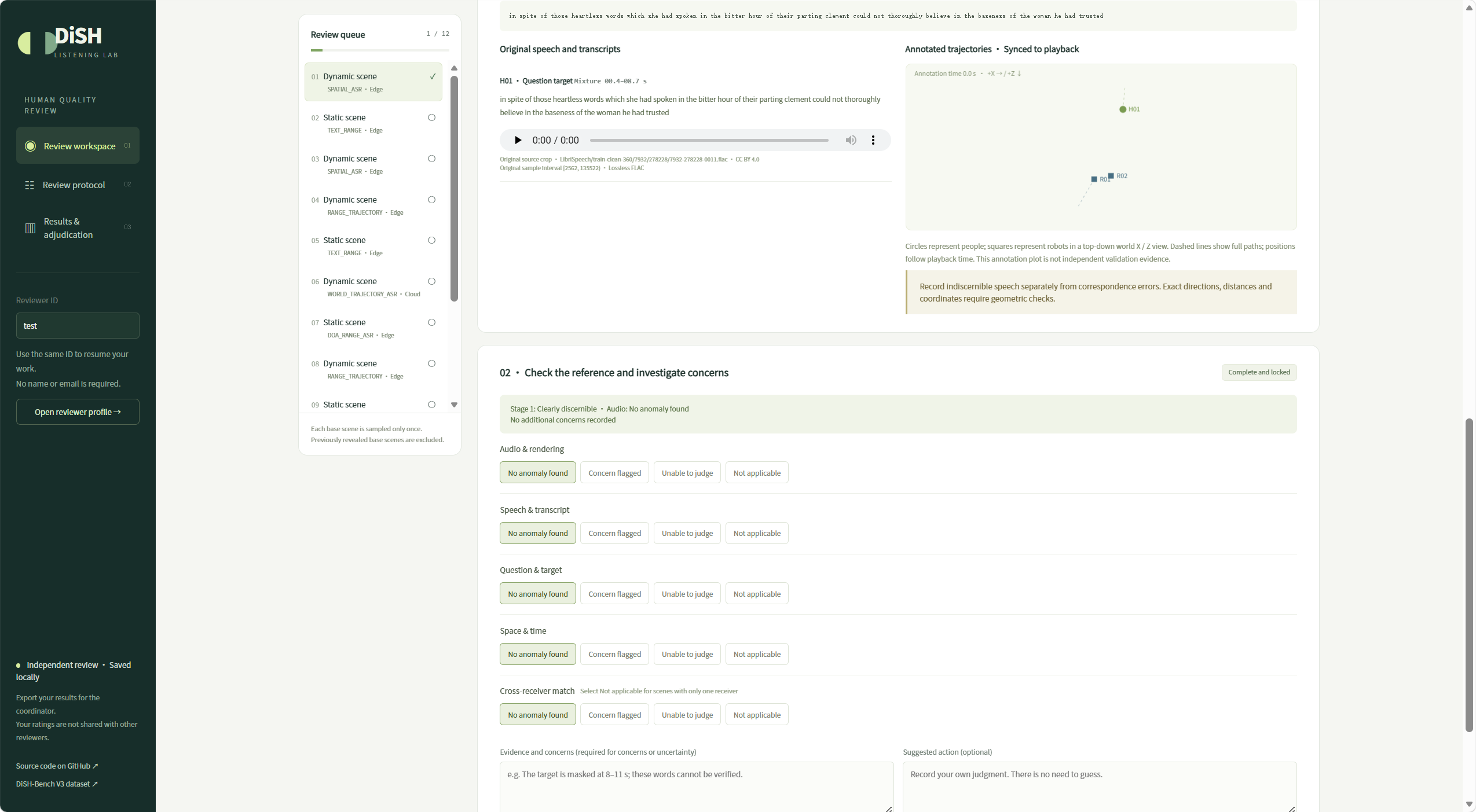}
\par\smallskip
{\small (b) Reference inspection and review form.}
\caption{The listening-review interface. (a) Reviewers configure the sampling
plan, navigate the review queue, inspect the question, and select a receiver
recording. (b) The reference-check view places source speech and transcripts
beside a playback-synchronized trajectory display. Separate controls record
concerns about audio rendering, speech--transcript correspondence,
question--target correspondence, space and time, and cross-receiver matching,
with fields for supporting evidence and suggested actions. The trajectory
display visualizes dataset annotations rather than independent geometric
validation.}
\label{fig:human-review-interface}
\end{figure*}

 A reproducible review record should
distinguish the presented recordings and query, the unedited model response,
the extracted answer, and the review decision.
Figure~\ref{fig:human-review-interface} shows the listening and reference-check
views of this interface.

\paragraph{Review interface.}
We implement a browser-based interface for two-stage dataset quality
review. It presents the original question and its permitted receiver
recordings, with waveform navigation, receiver switching at a shared
playback time, and timestamped annotations. Four-channel ACN/SN3D FOA is
decoded to binaural audio using a KEMAR first-order FIR filter bank.
Playback preserves channel relationships and uses a common gain across
receivers, with headphone and left/right calibration controls.
The first stage records target audibility and suspected rendering
anomalies before reference answers are revealed. The interface requires
at least 95\% playback coverage for each task-bound receiver before a
completed first-stage judgment can be locked; playback failures are
recorded separately. After locking, the second stage reveals the reference
answer, original source speech and transcripts, and trajectories synchronized
with playback. Reviewers separately assess rendering quality, speech--text
correspondence, question--target correspondence, spatial--temporal
continuity, and cross-receiver correspondence.

\paragraph{Sampling and review records.}
The interface supports reproducible sampling at the physical-realization
level, selecting one question per realization with fixed random seeds,
static/dynamic quotas, and room-wise sampling without repeated base scenes.
Exported plans fix the scene and question identities so that reviewers can
independently inspect the same material. Pilot and formal records are kept
separate. Ratings distinguish suspected annotation errors from masking
and other legitimate acoustic difficulty, and retain uncertainty,
non-applicability, comments, and affected time intervals.
Results are saved locally and explicitly exported for aggregation.
For shared samples, the analysis interface computes raw agreement and
nominal Krippendorff's $\alpha$ from independent ratings; adjudication is
stored separately without replacing those ratings. Pilot records, prior
reference exposure, and playback failures are excluded from formal
agreement analysis.

\paragraph{Listening audit.}
We collected 57 completed scene--question reviews covering
56 physical realizations, comprising 36 pilot and 21 formal reviews.
We report the two phases separately.
Before reference disclosure, 33 pilot and 18 formal samples
were rated clearly discernible; each subset contained two
partly discernible samples and one indiscernible sample.
Three pilot reviews flagged initial audio-quality concerns,
whereas none were flagged in the formal subset at this stage.
After reference disclosure, formal reviews flagged no concerns
about speech--text correspondence, question--target correspondence,
or spatial--temporal continuity.
Two formal records contained concerns about excessive background
noise and perceived overlap-related distortion, respectively.
One pilot record received concerns across all five review fields.
These concerns remain unadjudicated.

\paragraph{Scope.}
Each scene--question pair received a single review, precluding
inter-rater agreement estimates.
Formal playback logs covered only one receiver per question,
so the submitted cross-receiver ratings do not establish
cross-receiver consistency.
The findings provide preliminary evidence about perceptual
accessibility and potential quality issues in the reviewed subset.

\section{Qualitative Case Studies}
\label{app:case-traces}
\begin{center}\begin{minipage}{\linewidth}
\centering
\includegraphics[width=\columnwidth]{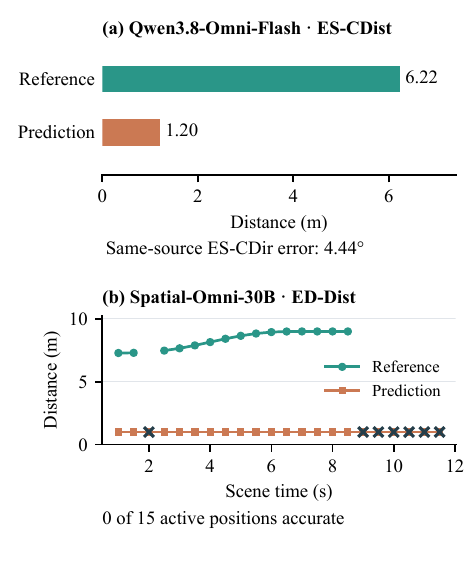}
\captionof{figure}{Direction can be accurate while distance is wrong (a); a well-formed track can ignore motion (b). Crosses indicate false speech activity.}
\label{fig:compact-cases}
\end{minipage}\end{center}

\begin{figure*}[p]
\centering
\includegraphics[width=\textwidth]{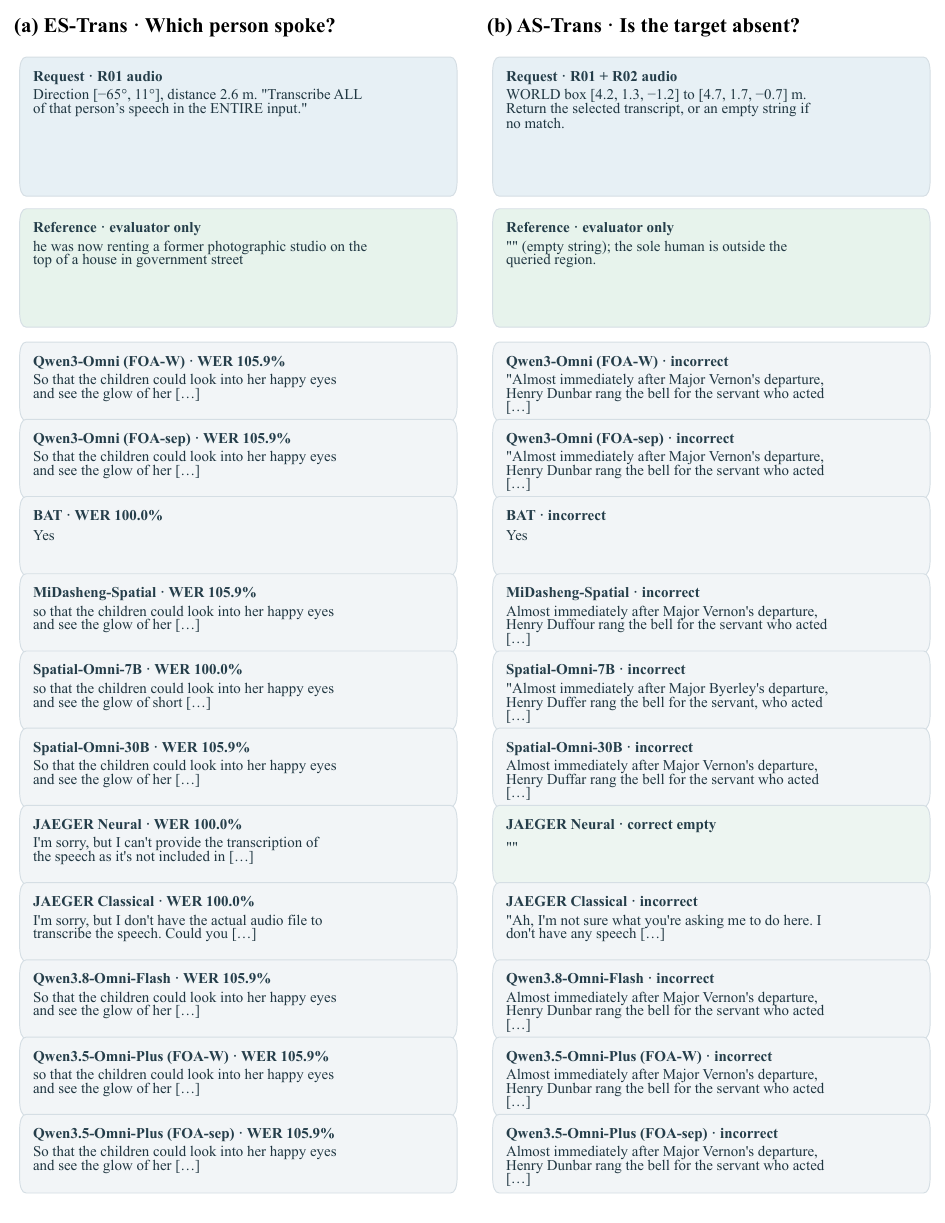}
\caption{The same two spatial queries across eleven direct configurations. Request boxes summarize the supplied cues; response boxes quote raw answers, with [\ldots] marking truncation. Scores use complete extracted answers. Samples: \texttt{00453216fc112c99fd735b2c} (a) and \texttt{00afedbe599a7a8dfd77cd0d} (b).}
\label{fig:shared-case-dialogues}
\end{figure*}

Figure~\ref{fig:shared-case-dialogues} compares all eleven direct configurations
on the same two queries. For ES-Trans, several models reproduce the wrong
human's sentence; for AS-Trans, several transcribe speech outside an empty
requested region. JAEGER Neural returns a correct empty answer in the latter
case. These examples distinguish intelligible recognition from spatial target
selection. Responses are frozen original outputs, excerpts are marked, and
scores use complete extracted answers; examples do not estimate prevalence.
\paragraph{MESH-agent on the shared queries.}
For the same ES-Trans example in
Figure~\ref{fig:shared-case-dialogues}, MESH-agent incurs 21 word edits
against a 17-word reference, corresponding to 123.53\% WER.
For the AS-Trans example with an empty target region, it returns a
nonempty answer, yielding 27 insertion errors and failing the correct-empty
criterion. Both answers pass the output-format check. These per-question
records establish answer-level failures but do not distinguish errors in
acoustic evidence acquisition, transcription, and final target selection;
that attribution requires the corresponding tool observations and decisions.

\end{document}